\documentclass{resources/aa}
\usepackage{graphicx}
\usepackage[varg]{txfonts}
\usepackage{physics}
\usepackage{amsmath, amssymb}
\usepackage{dsfont}
\usepackage{bbm}
\usepackage{xcolor}
\usepackage{ulem}
\usepackage{xpatch}
\usepackage{hyperref}
\usepackage{import}
\usepackage{makeidx}
\usepackage{changes}
\usepackage{hhline}
\usepackage{natbib}
\defcitealias{Cecil2024b}{Paper~I}
\usepackage{array}
\usepackage{makecell}
\usepackage{hhline}

\newcommand{\colout}[1]{\bgroup\markoverwith{\textcolor{#1}{\rule[.5ex]{2pt}{0.4pt}}}\ULon}

\makeatletter
\renewcommand*\aa@pageof{, page \thepage{} of \pageref*{LastPage}}
\newcommand{\pder}[2][]{\frac{\partial#1}{\partial#2}}
\makeatother

\renewcommand{\arraystretch}{1.5}

\makeindex

\begin{document}

\titlerunning{Inner disc edge with adaptive opacities}
\authorrunning{M. Cecil et al.}
\title{The role of detailed gas and dust opacities in shaping the evolution of the inner disc edge subject to episodic accretion } 
% \subtitle{I. }
%
% \subtitle{....}
%
\author{
 M.~Cecil\inst{1 \and 2}\thanks{Corresponding author; \texttt{cecil@mpia.de}},
 M.~Flock\inst{1},
 M. G.~Malygin\inst{1},
 R.~Kuiper\inst{3},
 P.~Sudarshan\inst{1 \and 2},
 A.~Ziampras\inst{1 \and 4},
 V. G. Elbakyan\inst{3}
}
\institute{
 Max Planck Institute for Astronomy (MPIA), Königstuhl 17, 69117 Heidelberg, Germany
 \and
 Fakultät für Physik und Astronomie, Universität Heidelberg, Im Neuenheimer Feld 226, 69120 Heidelberg, Germany
 \and
 Fakultät für Physik, Universität Duisburg-Essen, Lotharstraße 1, D-47057 Duisburg, Germany
 \and
 Ludwig-Maximilians-Universität München, Universitäts-Sternwarte, Scheinerstr 1, D-81679 München, Germany
}
%\today
\date{Received ....; accepted ....}

\abstract
{The transition in turbulence in the inner regions of protoplanetary discs and the closely connected dust sublimation front lead to periodic instability, manifesting as episodic accretion outbursts. For the corresponding interplay between heating and cooling, the opacity of the material needs to be treated carefully.
}
{We investigate the effects of different dust and gas opacity descriptions on the structure and evolution of the inner regions of protoplanetary discs. The influence on the episodic instability of the inner disc edge is hereby of central interest.  }
{Two-dimensional (2D) axisymmetric radiation hydrodynamic models are employed to simulate the evolution of the inner disc over the course of several thousand years. Our simulations greatly expand on previously published models by implementing detailed descriptions of the gas and dust opacities in terms of both their mean and frequency-dependent values. This allows us to also consider binned frequency-dependent irradiation from the central star.}
{The adaptive opacity description impacts the structure of the inner disc rim to a great extent, with the gas opacities' contribution having the most significant influence. The resulting effects include the shift in position of both the dust sublimation front and the dead zone inner edge (DZIE), a significantly altered temperature in the dust-free region and the manifestation of the equilibrium temperature degeneracy as a sharp temperature transition. The episodic instability due to the activation of the magneto-rotational instability (MRI) in the dead zone still occurs, but at lower inner disc densities. While the gas opacities set the initial conditions for the instability by determining the location of the DZIE, the evolution of the outburst itself is mainly governed by the dust opacities. The analysis of criteria for non-axisymmetric instabilities reveals possible breaking of the density peaks produced during the burst phase. However, due to the periodicity of the instability cycle, the DZIE itself may remain stable throughout the quiescent phases according to linear criteria applied to our axisymmetric models. }
{Although the thermal structure of the inner disc is crucially affected by different opacity descriptions, especially by the contribution of gas, the mechanism of the periodic instability of the DZIE remains active and is only marginally influenced by the gas opacities. The observational consequences of the severely altered temperatures may be significant and require further investigation. }


\maketitle

\section{Introduction} \label{sec:introduction}

A substantial amount of prior investigations has shown that the structure of the inner regions of protoplanetary discs is subject to extensive variability, originating from various processes and often manifesting as periodic outbursts of accretion onto the central star \citep{Lin1985, Kley1999, Wunsch2006, Vorobyov2006, Bae2013, Audard2014, Kadam2019, Steiner2021}. A compilation of observed variable young stellar objects is provided by \texttt{OYCAT} \citep{Pea2025}. The instabilities underlying the variability, as well as the potential luminosity feedback from the accretion burst, can have significant impacts on the disc's thermal and chemical structure \citep{Rab2017, Vorobyov2020, Laznevoi2025}. While most of the past studies focused on modelling observational signatures of outbursting young stellar objects such as FU Orionis \citep{Herbig1977}, our previous work \citep[][ hereafter \citetalias{Cecil2024b}]{Cecil2024b} investigated the instability of the inner disc as an intrinsic consequence of the dynamic and thermal evolution of the region around the dead zone inner edge (DZIE) over long periods of time. A particular focus was thereby laid on the potential impacts on the formation of planetesimals and planets in the inner disc.\par
The mechanism emerging in our radiation hydrodynamic simulations causes thermal instability (TI) by activating the magneto-rotational instability \citep[MRI,][]{Balbus1991} within the dead zone. While multiple previous studies have investigated such MRI-triggered outbursts \citep{Zhu2009, Kadam2020, Vorobyov2021, Cleaver2023, Das2025}, \citetalias{Cecil2024b} shows that they can naturally arise from the turbulence transition between the irradiation-dominated, gaseous inner disc and the optically thick outer domain. After activating the MRI beyond the DZIE, heating fronts travel into the dead zone, expanding the highly turbulent region and flushing large amounts of material onto the star. The burst phase is hereby substructured into multiple reflares of the instability. After the density of the inner regions has been lowered enough to prevent further MRI activation, the disc enters a quiescent phase, during which the inner dead zone is refilled with matter accreted from the outer regions. A new outburst cycle commences as soon as the combination of viscous heating and heat trapping raises the midplane temperature above the MRI activation threshold beyond the DZIE again. The temperatures reached during this burst process are not high enough to induce the `classical' TI by hydrogen ionisation \citep{Bell1993, Zhu2009b, Nayakshin2024, Elbakyan2024, Elbakyan2025}. \par
The type of accretion burst analysed in \citetalias{Cecil2024b} follows a limit-cycle and can be tracked on S-curves of thermal stability, as has been done in studies of cataclysmic variables \citep{Lasota2001, Hameury2020, Jordan2024}, but also for instabilities in protoplanetary discs \citep{Martin2011, Nayakshin2024}. Additionally, the burst cycle produces multiple pressure maxima, in which dust can potentially accumulate and grow \citep{Taki2016, Dullemond2018, Lee2022}. On the other hand, the pebble- and planet-trapping pressure bump at the DZIE \citep{Dzyurkevich2010, Flock2019, Chrenko2022} is periodically destroyed. \par
In \citetalias{Cecil2024b}, we used a simplified description of the opacities by fixing a single value for the gas contribution and two constant values for the mean and irradiation opacities, respectively, for the dust coefficients. However, in the context of instability cycles, the opacities govern the critical surface densities for activating ($\Sigma_\mathrm{crit}^\mathrm{max}$) and shutting down ($\Sigma_\mathrm{crit}^\mathrm{min}$) the underlying physical mechanism (MRI, in case of \citetalias{Cecil2024b}) \citep{Lodato2004, Wunsch2006, Nayakshin2024}. Furthermore, in regions where dust sublimation becomes relevant, the gas opacity can be dominant in describing heating and cooling as well as determining the optical thickness of the disc material and the shape of the inner dust rim \citep{Muzerolle2004, Isella2005, Isella2006, Zhang2011, Kuiper2012, Flock2016}. Therefore, we expanded on our previous models by implementing detailed descriptions of the dust and gas opacities in this work. While several versions of relevant gas opacity calculations have been presented in the literature \citep{Helling2000, Ferguson2005, Marigo2024}, the work of \cite{Malygin2014} provides frequency-dependent opacities in addition to the Planck and Rosseland means. This allows us to treat the irradiation in a frequency-dependent manner with contributions from the gas in addition to the effect of the frequency-dependence of dust opacities, which has already been shown to result in a more accurate description of radiative transport in irradiated discs \citep{Kuiper2013}. The mean gas opacities by \cite{Malygin2014} have also been used by \cite{Pavlyuchenkov2023} to show the multivalued equilibrium temperature solution leading to variable accretion, akin to the `classical' TI. \par
The density structure resulting from the burst cycle can be favourable for convergent migration of planetesimals, which has been proposed as a possible origin of the terrestrial planet distribution in the Solar System \citep{Ogihara2018, Broz2021}. However, the sharpness of the density features produced by the outburst mechanism calls their dynamic stability into question, especially regarding the Rossby wave instability \citep[RWI,][]{Lovelace1999}. The RWI can lead to the emergence of large-scale vortices \citep{Lyra2012, Bae2015, Flock2015}, which can affect the disc's dynamic structure and have been suggested as origins of dusty, non-axisymmetric structures in discs observed with ALMA \citep[e.g.,][]{Prez2018, Ziampras2025a}. Therefore, a careful investigation of the dynamic stability of spikes and bumps in the density structure during the burst and post-burst phases is warranted. \par
Our study intends to broaden our understanding of the inner disc's structure and dynamic evolution under the influence of periodic instability by carefully treating dust and gas opacities in both their mean and frequency-dependent form. Analogous to \citetalias{Cecil2024b}, we utilised two-dimensional, axisymmetric radiation hydrodynamic simulations, which are additionally complemented by the consideration of accretion luminosity feedback from the central host star. By subsequently activating the different descriptions in our simulations, we analyse their direct or combined impact. Additionally, we investigate the possibility of non-axisymmetric instability of the density features produced by the burst cycles. \par
The paper is structured as follows. In Sect.~\ref{sec:method}, we describe the physical and numerical configuration of our models. We present the results of the simulations in Sect.~\ref{sec:results} and discuss and analyse them further in Sect.~\ref{sec:discussion}. Our conclusions are summarised in Sect.~\ref{sec:conclusion}.

\section{Method}\label{sec:method}
The set-up and execution of the simulations conducted in this work follow the same strategy as presented in \citetalias{Cecil2024b}. Section~\ref{sec:meth_eqs} gives a brief overview of the governing equations and summarises the main parts of the modelling procedure. The subsequent sections are mainly concerned with the extension of the models of \citetalias{Cecil2024b} by implementing new physical prescriptions. Section~\ref{sec:opacities} presents the descriptions of gas and dust opacities, which are then included in the formulation of the frequency-dependent irradiation flux in Sect.~\ref{sec:meth_freqirr}. Section~\ref{sec:meth_viscosity} describes the implementation of the turbulent viscosity and Sect.~\ref{sec:numerica} addresses the numerical details. All simulations were conducted with the \texttt{PLUTO} code \citep{Mignone2007}, including the flux-limited diffusion (FLD) description of radiative transport presented in \cite{Flock2013}.
% The numerical consequences of the new opacity descriptions are presented in Sect. \ref{sec:meth_underrelax}.
\subsection{Governing equations} \label{sec:meth_eqs}
\begin{figure*}[t!]
    \centering
         \resizebox{\hsize}{!}{\includegraphics{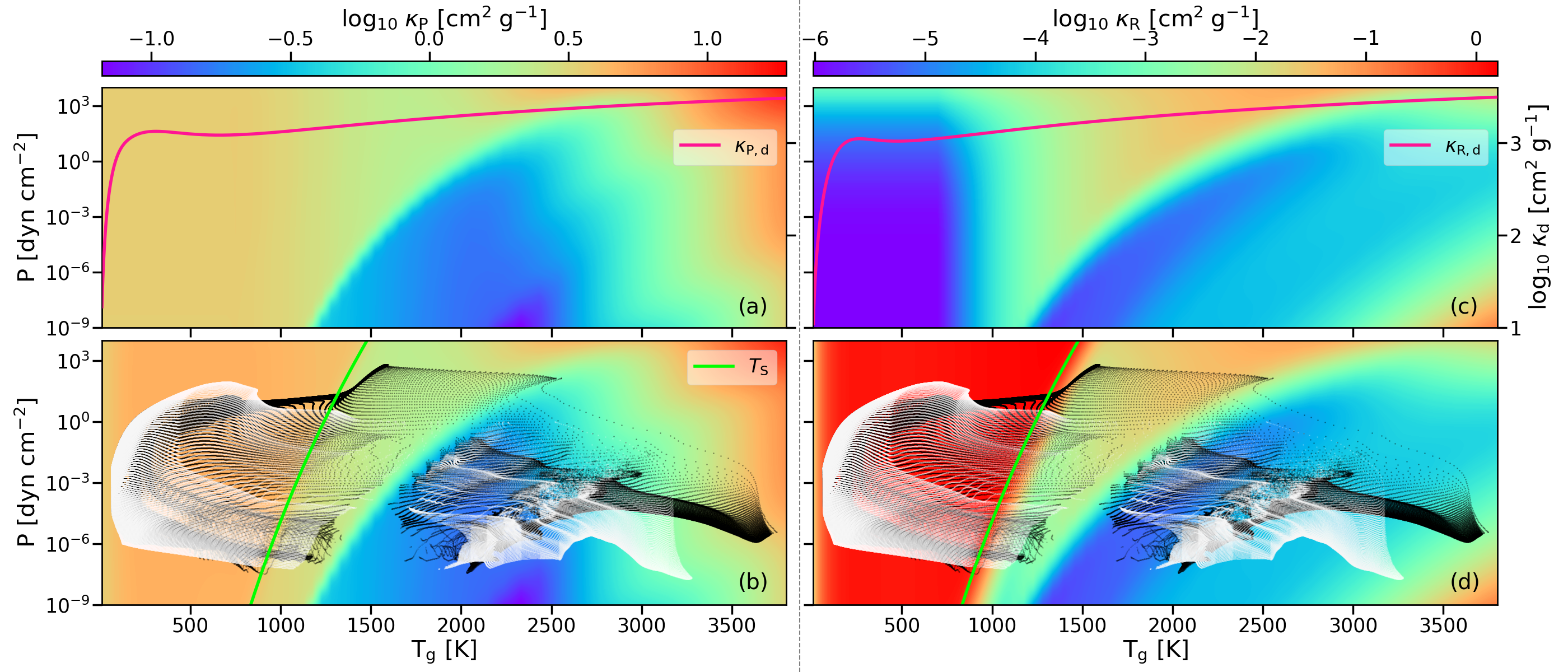}}
    \caption{Planck (panels a and b) and Rosseland (panels c and d) mean opacities used in our models. Panels (a) and (c) show maps of the temperature- and pressure-dependent gas mean opacities. The pink lines represent the dust mean opacities as functions of temperature. Panels (b) and (d) show the final effective mean opacities with the lime-coloured contour indicating the dust sublimation temperature, $T_\mathrm{S}$, for each pressure value. The white and black dots mark the temperature--pressure pairs that occur in our simulations, representing a snapshot of the quiescent and the outburst stage, respectively.}
    \label{fig:mean_opacs}
\end{figure*}
The initial model of the 2D axisymmetric disc in spherical co-ordinates $(r,\theta, \varphi)$ was constructed by solving the equations of vertical hydrostatic equilibrium in conjunction with the two radiative transport equations in the FLD approximation, including viscous heat dissipation as a source term,
\begin{align}
    &\frac{1}{\gamma -1} \, \pder[P_\mathrm{g}]{t}=-\kappa_\mathrm{P} \, \rho \, c \, (a_\mathrm{R} \, T_\mathrm{g}^4 - E_\mathrm{R}) - \nabla \cdot F_\mathrm{irr} + Q_\mathrm{acc} \; , \label{eq:rad1} \\
    &\pder[E_\mathrm{R}]{t} - \nabla \cdot \left ( \frac{c \, \lambda}{\kappa_\mathrm{R} \, \rho} \, \nabla E_\mathrm{R} \right ) = \kappa_\mathrm{P} \, \rho\, c \, (a_\mathrm{R} \, T_\mathrm{g}^4 - E_\mathrm{R}) \;,  \label{eq:rad2}
\end{align}
\noindent with the adiabatic factor $\gamma$, the gas thermal pressure $P_\mathrm{g}$, the mass density $\rho$, the gas temperature $T_\mathrm{g}$ (which we assume to be equivalent to the dust temperature), the radiation energy density $E_\mathrm{R}$, the azimuthal component of the viscous stress tensor $Q_\mathrm{acc}$, the flux-limiter function $\lambda$ \citep{Levermore1981}, the radiation constant $a_\mathrm{R}$ and the speed of light $c$. The description of the Planck and Rosseland mean opacities, $\kappa_\mathrm{P}$ and $\kappa_\mathrm{R}$, is given in Sect.~\ref{sec:opacities} and the form of the irradiation flux $F_\mathrm{irr}$ is presented in Sect.~\ref{sec:meth_freqirr}. \par
For creating the initial surface density profile $\Sigma$, which is then used to determine the two-dimensional density structure according to the hydrostatic equilibrium, we used,
\begin{equation} \label{eq:Sigma}
    \Sigma=\frac{\dot{M}_\mathrm{init}}{3\, \pi \, \nu_\mathrm{visc}} ~,
\end{equation} 
\noindent where $\dot{M}_\mathrm{init}$ is a radially constant mass flux through the disc and $\nu_\mathrm{visc}$ is the kinematic viscosity of a standard $\alpha$-disc model \citep{Shakura1973},
\begin{equation}
    \nu_\mathrm{visc}=\frac{\alpha c_\mathrm{s}^2}{\Omega}=\alpha  c_\mathrm{s}H \;,
\end{equation}
\noindent with $\alpha$ being the temperature-dependent stress-to-pressure ratio, the structure of which is described in Sect. \ref{sec:meth_viscosity}. We defined the speed of sound as $c_\mathrm{s}=\sqrt{\gamma P_\mathrm{g}/\rho}$ and the orbital frequency\footnote{We defined $r$ as the spherical radius. However, since $z/r$ is small in our models, we can assume $r\sim R$ with $R$ being the cylindrical radius.} as $\Omega=v_\varphi/r$, where $v_\varphi$ is the azimuthal component of the velocity field. The scale height $H$ is defined as $H=c_\mathrm{s}/\Omega$. \par
The hydrostatic initial model was then evolved in time by solving the coupled equations of continuity, motion and total energy,
\begin{align}
    &\pder[\rho]{t} \, + \nabla \cdot ( \rho \, \vec v ) = 0\;, \label{eq:cont}&& \\
    &\pder[\rho \, \vec v]{t} + \nabla \cdot (\rho \, \vec v  \, \vec v^T) + \nabla P_\mathrm{g} = -\rho \, \nabla \Phi + \nabla \cdot \vec \Pi \;, \label{eq:mot} &&
\end{align}
%\vspace{-0.7cm}
\begin{multline}
    \pder[E]{t} + \nabla \cdot [(E + P_\mathrm{g}) \, \vec v] = -\rho \, \vec v \cdot \nabla \Phi - \vec \Pi : \nabla \vec v  \\ 
    - \kappa_\mathrm{P} \, \rho \, c \, (a_\mathrm{R}  \, T_\mathrm{g}^4 - E_\mathrm{R}) - \nabla \cdot F_\mathrm{irr}  \;, \label{eq:ene} 
\end{multline}
\noindent together with the equations for radiative transport (Eqs. \ref{eq:rad1} and \ref{eq:rad2}), omitting the viscous heating term in Eq. \ref{eq:rad1} since it was already considered in the hydrodynamics equations through $\vec\Pi$. The velocity vector is denoted with $\vec v=(v_r, v_\theta, v_\mathrm{\varphi})$, $E$ is the total energy, $\Phi$ the gravitational potential and $\vec \Pi$ the viscous stress tensor. \par
As a closure relation, we used the ideal gas equation of state in both the hydrostatic and dynamic calculations, 
\begin{equation} 
    P_\mathrm{g}=\frac{\rho \, k_\mathrm{B} \, T_\mathrm{g}}{\mu_\mathrm{g} \, u} \; , \label{eq:state}
\end{equation}
\noindent with the mean molecular weight $\mu_\mathrm{g}=2.35$, the Boltzmann constant $k_\mathrm{B}$ and the atomic mass unit $u$. \par
As was the case in \citetalias{Cecil2024b}, due to the inclusion of viscous heat dissipation and the temperature-dependence of $\alpha$ in the establishment of $\Sigma$, we expect the initial hydrostatic model to be unstable to the TI by MRI activation in the dead zone. Consequently, the hydrodynamic simulations always start with a full TI accretion burst cycle. This initial phase can be regarded as a process of relaxation towards the beginning of the quiescent state, representing the disc structure equivalent to the state after subsequent burst cycles, as presented in \citetalias{Cecil2024b}. 

\subsection{Opacities} \label{sec:opacities}
The mean and frequency-dependent opacities for both gas and dust were interpolated from tabulated values in logarithmic space (assuming underlying power law relations between opacities and disc quantities) at each timestep and in every cell in the computational domain. The dependencies on disc properties and corresponding parameter spaces covered by the respective opacity tables are summarised in Table~\ref{tab:opacity_tables}. \par
For the dust component, we used the DIANA standard opacities \citep{Woitke2016}, calculated with \texttt{optool} \citep{Dominik2021}, with a minimum and maximum grain size of $a_\mathrm{min}=0.05\, \mu$m and $a_\mathrm{max}=10\, \mu$m, respectively, and a power law index of $a_\mathrm{pow}=-3.5$ \citep{Mathis1977} for the dust size distribution. The frequency-dependent dust opacities $\kappa_\mathrm{d}(\nu)$ from the \texttt{optool} output were then used to calculate the Planck and Rosseland means, $\kappa_\mathrm{P,d}$ and $\kappa_\mathrm{R,d}$. \par
The mean and frequency-dependent gas opacities were based on \cite{Malygin2014} \footnote{The mean gas opacity tables are only available at the CDS via anonymous ftp to cdsarc.u-strasbg.fr or via
\href{https://cdsarc.u-strasbg.fr/viz-bin/qcat?J/A+A/568/A91}{CDS-link}. Analogously, the frequency-dependent gas opacity tables are only available at the CDS as well, the link for which will be provided in the published version of this paper in A\&A.} and provided in separate tables. The Planck and Rosseland means, $\kappa_\mathrm{P,g}$ and $\kappa_\mathrm{R,g}$, were calculated on a grid of gas temperatures and pressures with ranges listed in Table~\ref{tab:opacity_tables}. Additionally, the table includes values of the gas density for each temperature--pressure pair, based on an ideal gas equation of state with varying mean molecular weight $\mu_\mathrm{g}$. The frequency-dependent gas opacities, $\kappa_\mathrm{g}(\nu)$, were tabulated on a slightly different, coarser grid of temperatures and pressures\footnote{In fact, the table included density values instead of the pressure grid. However, with the three grids ($T_\mathrm{g}, P_\mathrm{g}, \rho$) given in the gas mean opacity table, the underlying ideal gas equation of state is uniquely defined, which allows for an interpolation and simple transformation between the gas pressure and density values.} in addition to a range of wavelength bins\footnote{Consistency between the mean and frequency-dependent opacity tables has been confirmed by recalculating the mean opacities from the frequency-dependent ones and comparing them to the tabulated values. }. The opacity is given in the table as both a direct and harmonic average of the underlying fine-resolution spectrum within a wavelength bin. Our handling of the frequency-dependent opacities is elaborated on further in Sect.~\ref{sec:meth_freqirr}.\par

\begin{table}[t]
{\renewcommand{\arraystretch}{1.1}% tighter row height
\caption{Considered parameter ranges for the different opacity tables. }
\makegapedcells
\setcellgapes{1pt} % symmetric top/bottom gap inside every cell
\begin{tabular}{c|c|c|c}
\hhline{====}
\multicolumn{1}{l|}{} &
\makecell[c]{$T_\mathrm{g}$-range\\[-0.2em]{[}K{]}} &
\makecell[c]{$P_\mathrm{g}$-range\\[-0.2em]{[}$\mathrm{dyn}\;\mathrm{cm}^{-2}${]}} &
\makecell[c]{$\nu$-range {[}Hz{]}\\[-0.2em]($\lambda$-range {[}$\mu$m{]})} \\ \hline

$\kappa_\mathrm{P,R,d}$ &
\makecell[c]{$10^0 - 10^4$\\[-0.25em]\{500\}} &
\makecell[c]{--} &
\makecell[c]{--} \\ \hline

$\kappa_\mathrm{d}(\nu)$ &
\makecell[c]{--} &
\makecell[c]{--} &
\makecell[c]{$3\cdot 10^{13} - 3\cdot 10^{15}$\\[-0.25em]($0.1 - 10$)\\[-0.25em]\{50\}} \\ \hline

$\kappa_\mathrm{P,R,g}$ &
\makecell[c]{$7\cdot 10^2 - 10^4$\\[-0.25em]\{94\}\\[-0.25em]extrap.\ to $10^0$} &
\makecell[c]{$10^{-9} - 10^8$\\[-0.25em]\{126\}} &
\makecell[c]{--} \\ \hline

$\kappa_\mathrm{g}(\nu)$ &
\makecell[c]{$6\cdot 10^2 - 10^4$\\[-0.25em]\{60\}\\[-0.25em]extrap.\ to $10^0$} &
\makecell[c]{$3\cdot 10^{-13} - 10^{3}$\\[-0.25em]\{16\}} &
\makecell[c]{$3\cdot 10^{13} - 3\cdot 10^{15}$\\[-0.25em]($0.1 - 10$)\\[-0.25em]\{50\}} \\
\hline
\end{tabular}
\label{tab:opacity_tables}
}
\tablefoot{The number of sample points within the range of the according quantity is given in the curly brackets.}
\end{table}

Panels (a) and (c) of Fig. \ref{fig:mean_opacs} show the maps of the Planck and Rosseland mean gas opacities, respectively, in the temperature--pressure space relevant for the models in this work, together with the temperature-dependent dust mean opacities. While the gas pressure values occurring in our simulations lie well within the ranges of the grids, the temperatures can become significantly lower than 700\;K. In regions where this was the case, we extrapolated the opacities by adopting the values at the lower temperature grid boundary. This extrapolation has no effect in the optically thick regions of the disc near the midplane, where the diffusive radiative transport is primarily governed by the Rosseland mean opacities. Since the gas Rosseland mean is very small compared to its dust counterpart (panel c of Fig. \ref{fig:mean_opacs}) and dust will always be present at $T_\mathrm{g}<700 \, \mathrm{K}$, the effective Rosseland mean opacity will be dominated by the dust component. As a consequence, the extrapolation may only be relevant for the Planck mean opacities. We inferred from Fig. 4 in \cite{Malygin2014} that the Planck means remain approximately constant below 700\;K, consistent with our extrapolation, at least down to about 200\;K, according to \cite{Freedman2008}. Furthermore, the extrapolated Planck mean opacities only have an effect in intermediate disc layers above approximately two scaleheights (where the Rosseland means become subdominant with temperatures typically exceeding 200\;K) and below the hot disc atmosphere. Therefore, we do not expect this extrapolation to significantly affect the main findings of this work.\par
Panels (b) and (d) show the effective total opacities $\kappa_\mathrm{P,R}=\kappa_\mathrm{P,R,g}+f_\mathrm{D2G}\kappa_\mathrm{P,R,d}$ (with the dust-to-gas mass ratio $f_\mathrm{D2G}$) including the vaporisation of dust at a sublimation temperature $T_\mathrm{S}$ (green contour). We refer to Appendix \ref{app:d2g} for the parametrisations of $T_\mathrm{S}$ and $f_\mathrm{D2G}$.
Panels (b) and (d) also provide a visualisation of all combinations of gas temperature and pressure extracted from a representative simulation in states of both quiescence (white dots) and outburst (black dots). The noticeable gap in the coverage of the dots along the ridge of the Planck mean opacity transition is due to the equilibrium temperature degeneracy, the effects of which will be investigated in Sect. \ref{sec:res_freqdep}.

\begin{table*}[ht!]
{\renewcommand{\arraystretch}{1.35}
\caption{Model names and configurations}
\centering
\begin{tabular}{c|c|c|c|c|c}
\hhline{======}
Model                               & \makecell[c]{Dust mean opacity\\[-0.01em]{[}$\mathrm{cm}^2~\mathrm{g}^{-1}${]}} & \makecell[c]{Dust irradiation opacity\\[-0.01em]{[}$\mathrm{cm}^2~\mathrm{g}^{-1}${]}}  & \makecell[c]{Gas mean opacity\\[-0.01em]{[}$\mathrm{cm}^2~\mathrm{g}^{-1}${]}} & \makecell[c]{Gas irradiation opacity\\[-0.01em]{[}$\mathrm{cm}^2~\mathrm{g}^{-1}${]}}  & \makecell[c]{Accretion\\[-0.01em]luminosity} \\ \hline
$\texttt{MREF}^*$      & $\kappa_\mathrm{P}=\kappa_\mathrm{R}=700$                                                           & $\kappa_\mathrm{P}=1300$                                                                                    & $\kappa_\mathrm{P}=\kappa_\mathrm{R}=10^{-3}$                                                       & $\kappa_\mathrm{P}=10^{-3}$                                                                                                                                                    & no                                                            \\
\texttt{DUST}      & $\kappa_\mathrm{P,R}^\mathrm{DI}(T_\mathrm{g})$                                                                                              & $\kappa^\mathrm{DI}(\nu)$                                                                                                       & $\kappa_\mathrm{P}=\kappa_\mathrm{R}=10^{-3}$                                                       & $\kappa_\mathrm{P}=10^{-3}$                                                                                                                                                    & yes/no                                                        \\
\texttt{NOFREQIRR} & $\kappa_\mathrm{P,R}^\mathrm{DI}(T_\mathrm{g})$                                                                                                & $\kappa_\mathrm{P}^\mathrm{DI}(T_\star)$                                                                                    & $\kappa_\mathrm{P,R}^\mathrm{M14}(T_\mathrm{g}, P_\mathrm{g})$                                                                                 & $\kappa_\mathrm{P}^\mathrm{M14}(T_\star, P_\mathrm{g})$                                                                                                                                                    & no                                                            \\
\texttt{FULL}      & $\kappa_\mathrm{P,R}^\mathrm{DI}(T_\mathrm{g})$                                                                                               & $\kappa^\mathrm{DI}(\nu)$                                                                                                       & $\kappa_\mathrm{P,R}^\mathrm{M14}(T_\mathrm{g}, P_\mathrm{g})$                                                                                 & $\kappa^\mathrm{M14}(\nu,T_\mathrm{g}, P_\mathrm{g})$                                                                                                                                                             & yes/no  \\                                                      
\hline                                                          
\end{tabular}
\label{tab:models}
\tablefoot{The superscript DI indicates opacities according to the DIANA standards, while M14 symbolises gas opacities from \cite{Malygin2014}. }
}
\end{table*}

\subsection{Irradiation flux} \label{sec:meth_freqirr}
\begin{figure}[t]
    \centering
         \resizebox{\hsize}{!}{\includegraphics{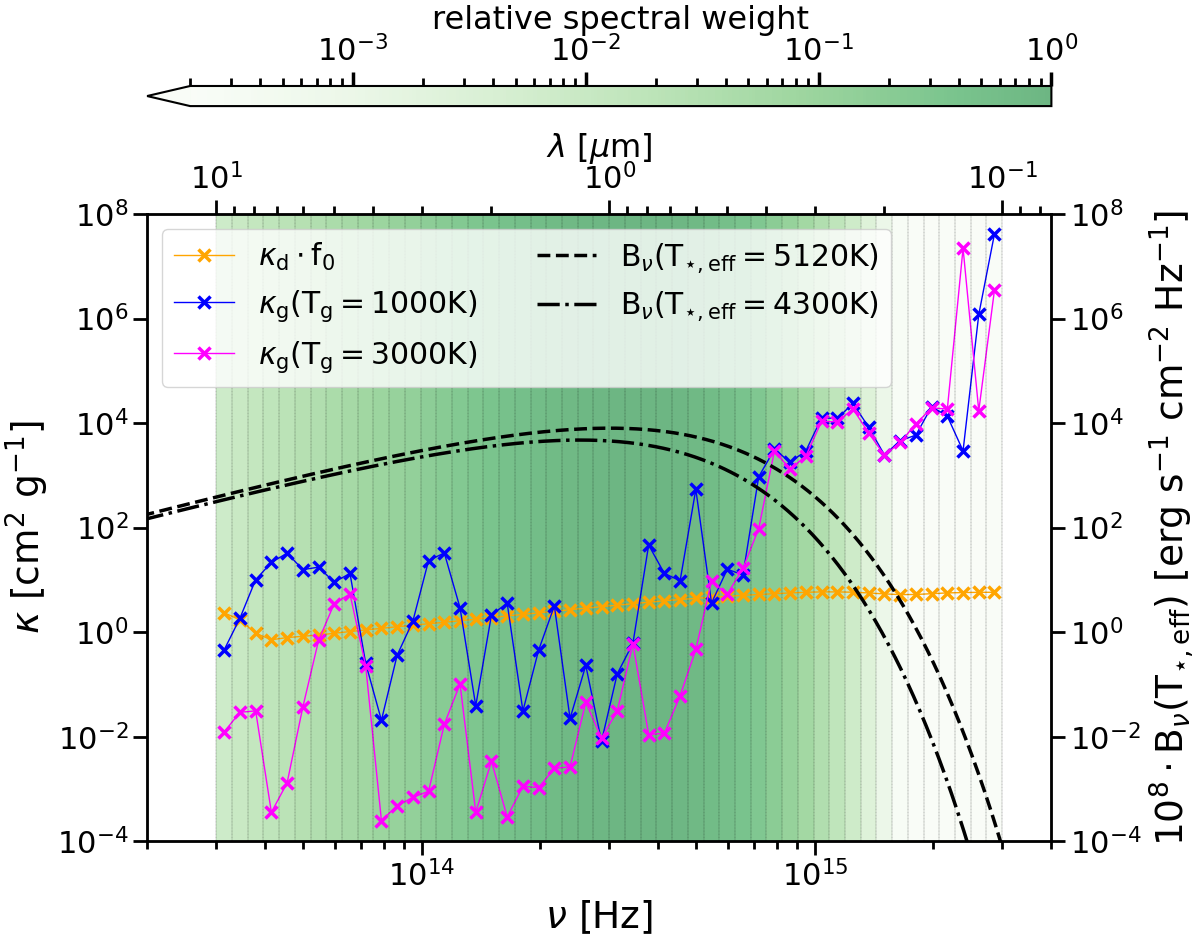}}
    \caption{
    Frequency-dependent gas and dust opacities in relation to the irradiating black body spectrum of the central star. The black lines indicate the Planck function of the irradiation, where the dash-dotted line represents the star during the quiescent phase (with negligible contribution from the accretion luminosity), whereas the dashed line shows the irradiating spectrum that includes the effect of the maximum accretion luminosity occurring in our models. The separation of the frequency space into the 50 bins is indicated by the thin vertical lines, where each bin is coloured according to its relative spectral weight. The orange crosses mark the dust opacities in their respective bins, multiplied by the maximum dust-to-gas mass ratio $f_0$. The gas opacities in each bin have been evaluated for a density of $10^{-10}\,\mathrm{g\,cm^{-3}}$ and are shown for two representative temperatures: 1000\,K (blue crosses) and 3000\,K (magenta crosses). 
    }
    \label{fig:B_star}
\end{figure}
The method used in our models to efficiently describe the frequency dependence of the stellar irradiation was based on \cite{Kuiper2010}. In the following, we review the method's specific implementation in our models. We hereby expand upon the description used by \cite{Sudarshan2025} by including the frequency-dependent gas opacities. \par
We calculated the total irradiation flux at every radius with (we refer to Appendix \ref{app:F_irr} for a derivation), 
\begin{equation} \label{eq:F_irr_freqdep}
    F_\mathrm{irr}(r)= \left(\frac{R_\star}{r}\right)^2 \sigma_\mathrm{SB}T_{\star,\mathrm{eff}}^4 \sum_{i=1}^n w_ie^{-\tau_\mathrm{rad}(\nu_i,r)} ~~~.
\end{equation}
\noindent with the stellar radius $R_\star$, the Stefan-Boltzmann constant $\sigma_\mathrm{SB}$, the number of frequency bins $n$, the central frequency of a bin $\nu_i$, its spectral weight $w_i$ and its radial optical depth $\tau_\mathrm{rad}(\nu_i,r)$. The effective stellar temperature $T_{\star,\mathrm{eff}}$ takes the contribution of accretion shock luminosity (following the description given in Appendix \ref{sec:meth_acclum}) into account. The spectral region considered for the frequency-dependent opacities extends from 0.1\,$\mu$m to 10\,$\mu$m and is separated into 50 bins as a compromise between computational efficiency and accuracy. Since the absorption of irradiation is typically governed by the Planck means, we chose the direct averages within the frequency bins for the gas opacities. A visualisation of the binned frequency-dependent opacities for both gas and dust, in relation to the irradiating stellar spectrum, is given in Fig. \ref{fig:B_star}. The temperatures of the displayed blackbody spectra include contributions from both the stellar and accretion shock luminosity. The colour coding of the bins corresponds to their relative spectral weight $w_i/\mathrm{max}(w_i)$. Following the definition of $w_i$ in Appendix \ref{app:F_irr}, the bin with the largest weight (i.e.~carrying most of the energy) is located at the maximum of $B_\nu(\nu, T_{\star, \mathrm{eff}})$, which shifts only slightly under the influence of the maximum accretion luminosity. The wavelength range beyond 10 $\mu$m, which we do not consider in our irradiation description, only incorporates $\sim \hspace{-0.08cm}0.2\%$ of the total irradiation energy and can, therefore, be neglected. The gas opacities are displayed for two different temperatures, with the density being kept constant. When considering the effect of the dust-to-gas mass ratio, the contribution of the gas to the total opacity can significantly exceed the impact of the dust, especially at small wavelengths and low temperatures. \par
Without frequency-dependent irradiation, we can set $n=1$ and $w_i=w=1$ in Eq. \ref{eq:F_irr_freqdep}, $\tau_\mathrm{rad}(\nu,r)$ becomes the total radial optical depth $\tau_\mathrm{rad}(r)$ and we arrive at the usual expression for the grey irradiation flux,
\begin{equation}\label{eq:F_irr_nofreqirr}
    F_*(r)=\left ( \frac{R_*}{r} \right) ^2\, \sigma_\mathrm{SB}\; T_{\star, \mathrm{eff}}^4\; e^{-\tau_\mathrm{rad}(r)} \; .
\end{equation}

\subsection{Viscosity} \label{sec:meth_viscosity}

For most models in this work, we used the same description for the temperature-dependent viscous $\alpha$ parameter as in \citetalias{Cecil2024b}, 
\begin{equation}
    \alpha=(\alpha_\mathrm{MRI}-\alpha_\mathrm{DZ}) \frac{1}{2} \left [1 - \mathrm{tanh}\left ( \frac{T_\mathrm{MRI}-T_\mathrm{g}}{T_\Delta}\right)\right]+\alpha_\mathrm{DZ} \; , \label{eq_alpha}
\end{equation}
\noindent where $\alpha_\mathrm{MRI}$ and $\alpha_\mathrm{DZ}$ describe the viscosity of the MRI active and dead zone regions, respectively, and $T_\mathrm{MRI}$ represents the threshold temperature around which the MRI becomes active. Following \cite{Flock2019} and \citetalias{Cecil2024b} and the motivations given therein, we set $\alpha_\mathrm{MRI}=0.1$,  $\alpha_\mathrm{DZ}=10^{-3}$ and $T_\mathrm{MRI}=900\,\mathrm{K}$. While we implemented a narrow transition between $\alpha_\mathrm{MRI}$ and $\alpha_\mathrm{DZ}$ with a smoothing parameter $T_\Delta=25 \,\mathrm{K}$ for the models in \citetalias{Cecil2024b}, we expanded the transition range for the new models of this work to $T_\Delta=70 \,\mathrm{K}$. The influence of this parameter, together with the effects of a smaller value of $\alpha_\mathrm{DZ}$, is investigated in Appendix \ref{app:alphathings}. \par
Previous studies have implied that the effective viscosity does not instantaneously respond to changes in the disc properties \citep[e.g.][]{Hirose2009, Flock2017a, Ross2017, Held2022}{}{}. Instead, there may be a significant delay in $\alpha$ adapting to alterations of the thermal and ionisation structure. We analyse the effects of such a delay by adapting Eq.~\ref{eq_alpha} accordingly and incorporating this description in additional models. We present the modelling procedure and its consequences in Appendix \ref{app:alphadelay}.

\subsection{Numerical considerations} \label{sec:numerica}
The majority of the numerical parameters and boundary conditions were directly adopted from \citetalias{Cecil2024b}. Here we briefly present the relevant differences. \par
To allow for efficient cooling by radiation, the vertical domain has to be large enough to capture the transition from the optically thick disc to the optically thin atmosphere at every radius. Additionally, the polar boundary conditions should not inhibit the cooling process. In ensuring that, taking into account the new opacity descriptions of the models of this work, the polar domain was expanded to $N_\theta=192$ linearly spaced cells and covered a range in $\theta$ of $\pi/2 \pm 0.22$, in contrast to $\pi/2 \pm 0.15$ considered in our previous models. The radial domain remained separated into $N_r=2048$ logarithmically spaced cells and extended from $r_\mathrm{in}=0.05\,\mathrm{AU}$ to $r_\mathrm{out}=10 \; \mathrm{AU}$.
As a polar boundary condition for the temperature, we set,
\begin{equation}
    T_\mathrm{0}=0.6 \left (\frac{R_*}{2r}\right )^{1/2}\, T_* \; ,
\end{equation}
\noindent which is very similar to the condition used in \citetalias{Cecil2024b}. It has been ensured that the resulting values are smaller than the typical temperatures occurring near the boundaries in our models, therefore enabling the disc to cool efficiently. \par
The sensitive temperature- and pressure-dependence of both the mean and frequency-dependent opacities necessitated the application of an under-relaxation scheme to the solver algorithm to maintain numerical stability. To exclude any impact on the onset and evolution of the episodic accretion events, the under-relaxation scheme was only applied in the optically thin disc atmosphere, where numerical instabilities are most prevalent. The detailed strategy and consequences of this method are explored in Appendix \ref{app:underrelax}.

\section{Results}\label{sec:results}
We conducted a multitude of two-dimensional, radiation hydrodynamic simulations with different combinations of the newly implemented physical effects. Table \ref{tab:models} lists the names and configurations of the various models presented in this section, including $\texttt{MREF}^*$ from \citetalias{Cecil2024b} as a reference. While the model \texttt{DUST} only considers the effects of the DIANA standard dust opacities for both the mean values and the frequency-dependent irradiation, \texttt{NOFREQIRR} combines the new descriptions for mean opacities of both the dust and the gas, but the irradiation is treated as frequency-integrated (i.e. grey irradiation), using the Planck means as absorption coefficients. The \texttt{FULL} model incorporates the new mean opacities and frequency-dependent irradiation, taking into account the contributions of both gas and dust. Additionally, the models \texttt{DUST} and \texttt{FULL} have been simulated with and without the effect of accretion luminosity feedback. 
The values of the remaining physical and numerical parameters used in all simulations are listed in Table \ref{tab:MREF}.

\begin{table}[]
{\renewcommand{\arraystretch}{1.1}
\caption{Model parameters.}
\begin{tabular}{lr|lr}
\hhline{====}
$M_*$ {[}$\mathrm{M}_\odot${]}                           & 1.0                & $T_\Delta$ [K]                                                    & 70              \\
$R_*$ {[}$\mathrm{R}_\odot${]}                           & 2.6                & $f_0$                                                    & $10^{-3}$               \\
$T_*$ {[}K{]}                                            & 4300               & $r_\mathrm{in}$ {[}AU{]}                                                  & 0.05         \\
$\alpha_\mathrm{MRI}$                                    & 0.1                & $r_\mathrm{out}$ {[}AU{]}                                                 & 10              \\
$\alpha_\mathrm{DZ}$                                     & $10^{-3}$          & $\theta$ {[}rad{]}                                                        & $\pi /2 \pm 0.22$                \\
$\dot{M}_\mathrm{init}$ {[}$M_\odot \, \mathrm{yr}^{-1}${]} & $3.6\cdot 10^{-9}$ & $N_\mathrm{r}$                                                            & 2048 \\
$T_\mathrm{MRI}$ {[}K{]}                                 & 900                & $N_\theta$                                                                & 192              \\

\hline
\end{tabular}
\tablefoot{These parameters are implemented in the models \texttt{DUST, NOFREQIRR, FULL} and \texttt{FULL\_DEL}. The differences to $\texttt{MREF}^*$ are restricted to the configuration of the polar domain and $T_\Delta$ and are laid out in Sect.~\ref{sec:method}. }
\label{tab:MREF}
}
\end{table}
\subsection{Influence of dust opacities} \label{sec:res_dust}
\begin{figure}[t]
    \centering
         \resizebox{\hsize}{!}{\includegraphics{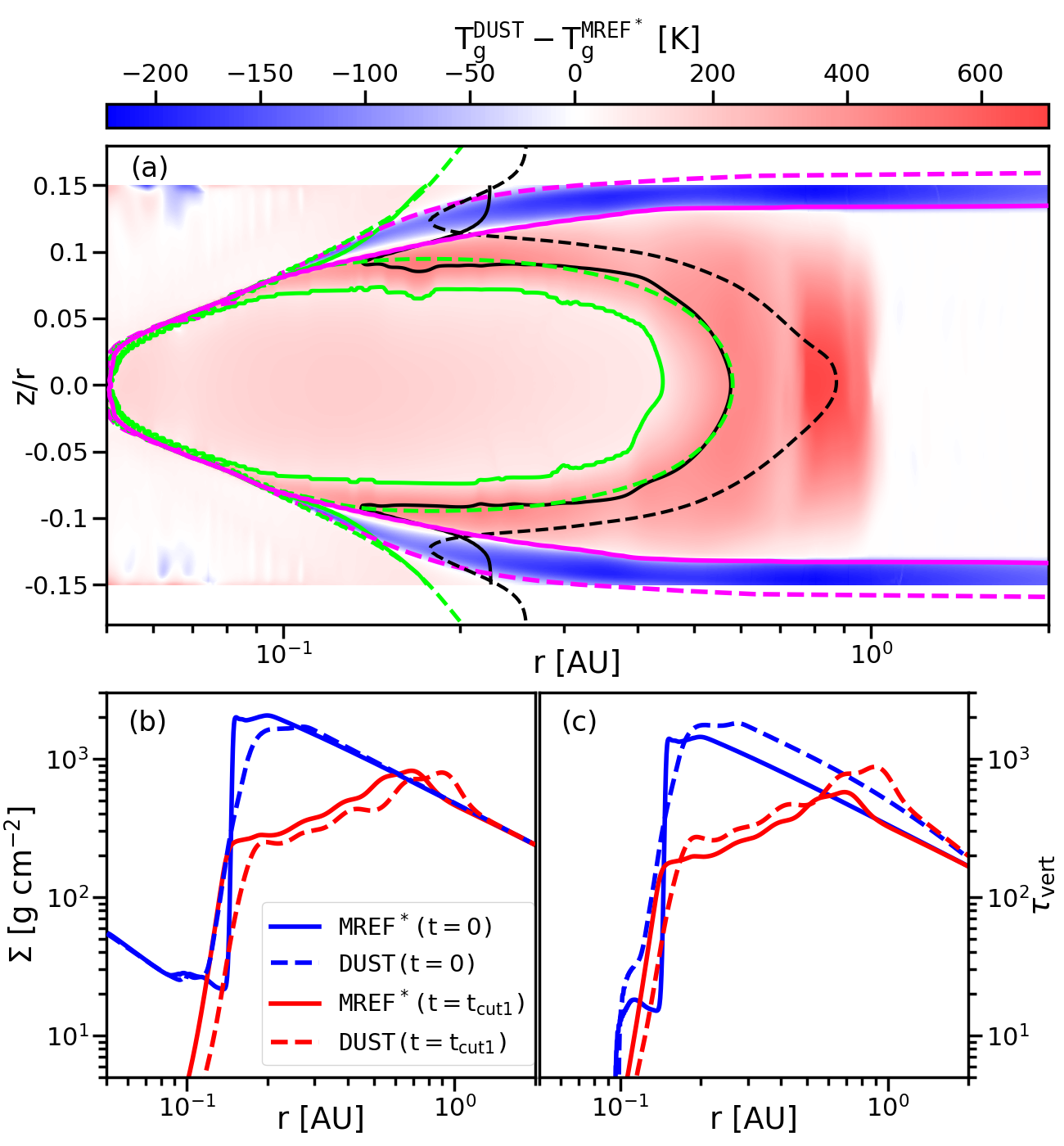}}
    \caption{Comparison between the models $\texttt{MREF}^*$ and \texttt{DUST}. Panel (a) shows a map of the temperature difference between the two models at a stage when the MRI active region has reached its largest extent during a burst. The black, green and magenta contour lines represent the MRI transition, the dust sublimation front and the $\tau_\mathrm{rad}=1$ surface, respectively, where the dashed lines correspond to \texttt{DUST} and the solid lines to $\texttt{MREF}^*$. Panels (b) and (c) show the surface density and the total vertical optical depth, respectively, of both models at two different stages: the initial hydrostatic structure (blue) and the state after the initial burst, when most of the density bumps have diffused (red). }
    \label{fig:MREF_DUST_diff}
\end{figure}

In order to isolate the effect of the new description of dust opacities via the DIANA standards, we analyse the initial outbursts occurring in the models $\texttt{MREF}^*$ and \texttt{DUST}, starting from essentially the same initial hydrostatic structure. 
Fig. \ref{fig:MREF_DUST_diff} shows a comparison between these two models in the initial configuration, the outburst stage and the post-burst structure. The small difference in the gradient of the initial surface density profiles at the DZIE, recognisable in panel (b), is due to the discrepancy between the values of $T_\Delta$ of the two models. The vertical optical depths of the initial models shown in panel (c) reveal that the DIANA opacity description results in a larger optical thickness of the disc compared to the constant opacity values in $\texttt{MREF}^*$. For instance, at a radius of 1\,AU, $\tau_\mathrm{vert}$ differs by a factor of 1.5 between \texttt{DUST} and $\texttt{MREF}^*$. Since the surface density is equal outside a radius of $\sim$$ 0.2 \,\mathrm{AU}$, the heat produced by viscous accretion near the midplane can get trapped more efficiently in the case of \texttt{DUST}. This has a variety of consequences for the evolution of the TI cycle. Comparing the MRI activation fronts in panel (a) (black contours) reveals that a considerably larger portion of the dead zone is heated up and made MRI active in the \texttt{DUST} model. The heating front, launched after activating the MRI near the DZIE, is able to travel further outwards compared to the optically thinner disc in $\texttt{MREF}^*$, where radiative cooling is more efficient. Consequently, the outermost density- and pressure bump emerging from the burst cycle (i.e. the peak in the surface density in the post-burst state) lies at a larger radius in the \texttt{DUST} model. Furthermore, the MRI active region becomes hotter, allowing for a larger fraction of the dust content to be sublimated. Hence, the upper equilibrium branch of the S-curve describing the instability cycle shifts to higher temperatures (as explored further in Sect.~\ref{sec:dis_Scurve}). \par
The magenta coloured lines in panel (a) of Fig.~\ref{fig:MREF_DUST_diff} mark the $\tau_\mathrm{rad}=1$ surfaces for the stellar irradiation. In the case of \texttt{DUST}, $\tau_\mathrm{rad}$ has to be understood as the optical depth considering the total irradiation flux, $\tau_\mathrm{rad,tot}$. Hereby, $\tau_\mathrm{rad,tot}=1$ has been calculated as the radius at each height at which $F_\mathrm{irr}(r)$ (Eq. \ref{eq:F_irr_freqdep}) has decreased to $F_\mathrm{irr}(R_\star)e^{-1}$. The surface lies at larger heights compared to $\texttt{MREF}^*$, creating the drop in relative temperature between the two magenta lines due to the stellar irradiation being absorbed sooner behind the dust sublimation front in \texttt{DUST}. \par
Another consequence of the larger dust opacities in \texttt{DUST} becomes discernible in the comparison of the post-burst states. Since the MRI can be kept active even at lower surface densities as compared to $\texttt{MREF}^*$, the TI cycle in \texttt{DUST} not only engulfs a greater part of the dead zone, but also removes more mass from the impacted regions. As a consequence, although the overall shape of the surface density profile in the post-burst state is the same in both models (including the positive gradient in the burst region), the absolute values of $\Sigma$ in \texttt{DUST} are lower. Despite the smaller surface density, the vertical optical depth in the post-burst state is still larger. This can be understood by considering that the minimum surface density necessary to keep the MRI active ($\Sigma_\mathrm{crit}^\mathrm{min}$) depends on the balance between heating and cooling efficiency, both of which depend on the surface density. In the \texttt{DUST} model, the more potent heat trapping near the midplane has to be compensated by a greater diminishment of the surface density to enable effective cooling. However, a smaller surface density also decreases the viscous heating efficiency, which results in the equilibrium at $\Sigma_\mathrm{crit}^\mathrm{min}$ allowing for a larger vertical optical depth. \par
Sections \ref{sec:res_gasopac} and \ref{sec:dis_Scurve} will show that the evolution of the disk during burst cycles is mainly determined by the dust opacities, while the contribution of gas is only marginal. Therefore, the analysis of the onset, progression and consequences of the outburst under the influence of the DIANA opacities will be conducted in the context of the \texttt{FULL} model in the following sections.

\subsection{Influence of gas opacities} \label{sec:res_gasopac}

\begin{figure}[t]
    \centering
         \resizebox{\hsize}{!}{\includegraphics{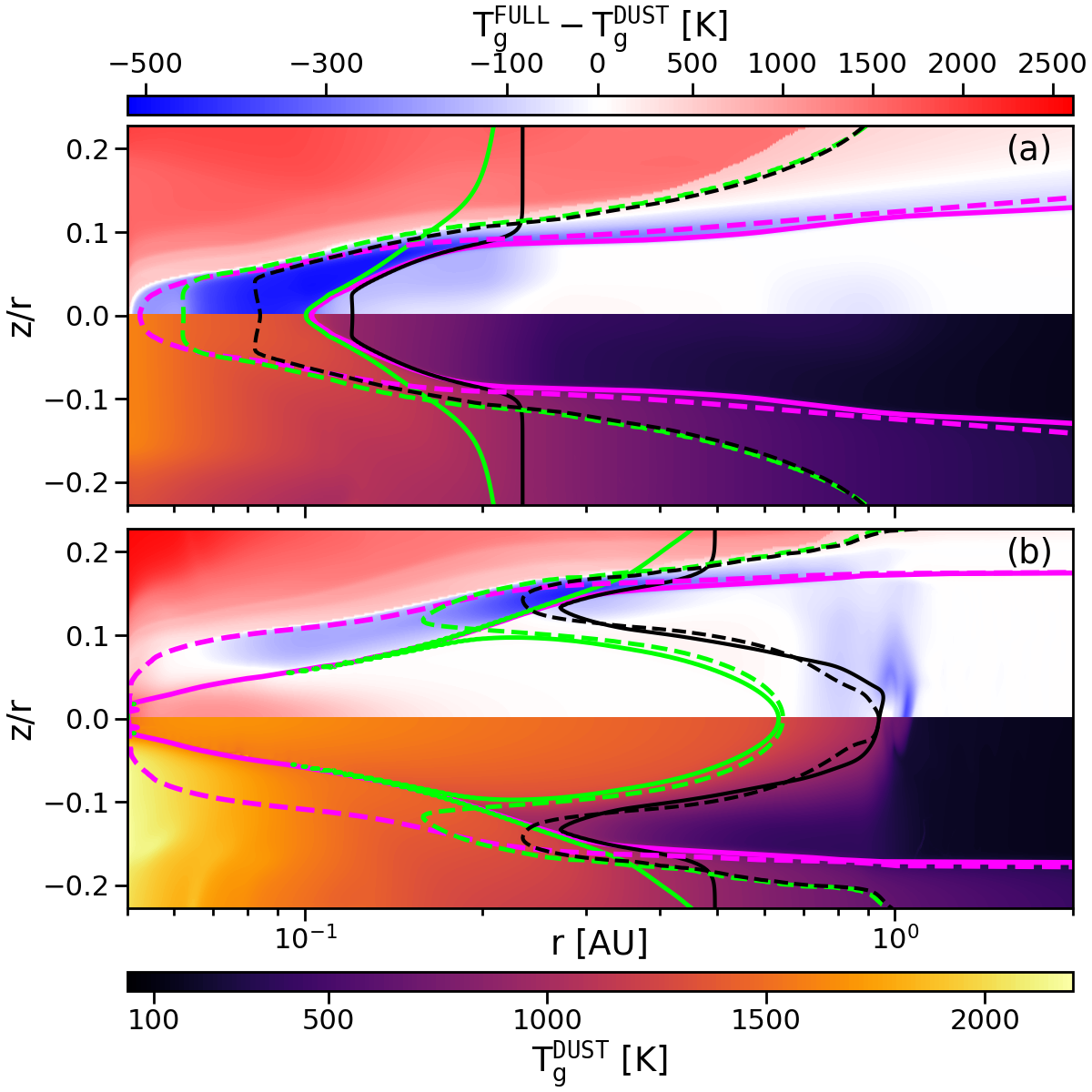}}
    \caption{Differences between the models \texttt{DUST} and \texttt{FULL} during quiescence (panel a) and outburst (panel b). The coloured contour lines represent the same transitions as in Fig.~\ref{fig:MREF_DUST_diff}, with the solid lines corresponding to \texttt{DUST} and the dashed lines to \texttt{FULL}. The upper halves of the panels show the differences in temperature between the two models, while the lower halves depict the absolute temperature of \texttt{DUST}. }
    \label{fig:old_new_diff}
\end{figure}

The \texttt{FULL} model incorporates the mean and frequency-dependent gas opacities, as calculated by \cite{Malygin2014}, in addition to the DIANA standard dust opacities. We start the analysis of their combined effects by comparing \texttt{FULL} to the previously investigated \texttt{DUST} model. 
Figure \ref{fig:old_new_diff} displays the discrepancies between these two configurations for both the quiescent (panel a) and the outburst phase (panel b). Additionally, the lower hemispheres of the two depicted snapshots show the temperature map of \texttt{DUST}. In both models, the accretion shock luminosity feedback is included. \par
The structure of the quiescent inner disc shows several remarkable differences. The temperature in the optically thin, gaseous atmosphere is significantly higher (by up to $2000\,\mathrm{K}$) in the case of the \texttt{FULL} model. While we assumed a constant value of $10^{-3}\,\mathrm{cm^2\,g^{-1}}$ for both the irradiation and Planck mean gas opacity in the \texttt{DUST} model, the corresponding opacities in \texttt{FULL} adapt freely to the conditions in the atmosphere, typically exceeding $0.1\,\mathrm{cm^2\,g^{-1}}$ (see Fig.~\ref{fig:mean_opacs}), leading to a higher temperature in the equilibrium between heating and cooling. For the definition and evaluation of the equilibrium temperature, we refer to Appendix~\ref{app:Teq} and Sect.~\ref{sec:res_freqdep}. \par
The shapes and positions of the MRI transition, the dust sublimation front and the $\tau_\mathrm{rad,tot}=1$ surface are strongly altered by the adaptive gas opacities. Due to the typically higher irradiation opacity of the \texttt{FULL} model compared to \texttt{DUST}, the stellar radiation is absorbed much sooner in the vicinity of the midplane, shifting the  $\tau_\mathrm{rad,tot}=1$ surface significantly closer towards the star. Consequently, the midplane temperature decreases below the dust sublimation and MRI activation thresholds much sooner, placing these transitions at the midplane at smaller radii as well. As a result, there is only a very small region ($<\hspace{-0.3em}0.015\, \mathrm{AU}$) at the disc's midplane that can be considered as dust-free, assuming that the magnetic truncation radius lies approximately at the inner boundary. This is the case even though we neglected the attenuation of the stellar irradiation between the stellar surface and our computational domain. \par
Similar to what has been explored in Fig.~\ref{fig:MREF_DUST_diff}, the region between the two $\tau_\mathrm{rad,tot}=1$ surfaces is significantly cooler in the \texttt{FULL} model. At a height of $z/r\sim\pm 0.1$, the gaseous atmosphere is optically thin enough in both models such that the $\tau_\mathrm{rad,tot}=1$ surfaces approximately coincide again outside of $0.2\, \mathrm{AU}$, where the optical depth is mainly determined by the dust. \par
While the shapes of the transitional surfaces show a `pointy' feature at the midplane in the \texttt{DUST} model, they form a more rounded structure and become very shallow at a height of $z/r\sim\pm 0.05$ in the case of \texttt{FULL}. The MRI activation front and the inner dust rim become vertical just outside $0.2\, \mathrm{AU}$ in \texttt{DUST}, whereas they approach the polar boundaries only slowly in \texttt{FULL}, maintaining a hot, gaseous atmosphere at much larger radii. \par

Analogous to panel (a) of Fig.~\ref{fig:MREF_DUST_diff}, panel (b) of Fig.~\ref{fig:old_new_diff} compares the structure of \texttt{DUST} and \texttt{FULL} at the respective times when the MRI active region has reached its greatest extent during an accretion event. The depicted outbursts are again part of the initial cycle, starting from the same hydrostatic initial model. The comparison of the MRI transitions (black contours) and dust sublimation fronts (green lines) indicates that the instability affects almost exactly the same region of the dead zone in both models, hinting at the dominance of the dust opacity, rather than the gas contribution, during the burst. Within the MRI active region near the midplane, the temperatures only start to diverge very close to the star, where the material becomes hot enough to sublimate the dust content to an extent at which the gas opacity (especially $\kappa_\mathrm{R,g}$) becomes influential again. The similarities in the optically thick parts of the disc during the outburst, as well as the slight differences, manifest themselves in the shapes and positions of the S-curves, which will be investigated further in Sect.~\ref{sec:dis_Scurve}. \par
In front of the inner dust rim, the $\tau_\mathrm{rad,tot}=1$ lines differ between the two models, similar to what was observed in the quiescent phase. Due to the large amounts of material being accreted inwards during the outburst cycle, the midplane becomes optically thick enough to allow for the $\tau_\mathrm{rad,tot}=1$ surfaces to reach the inner boundary in both models.

\begin{figure*}[]
    \centering
         \resizebox{\hsize}{!}{\includegraphics{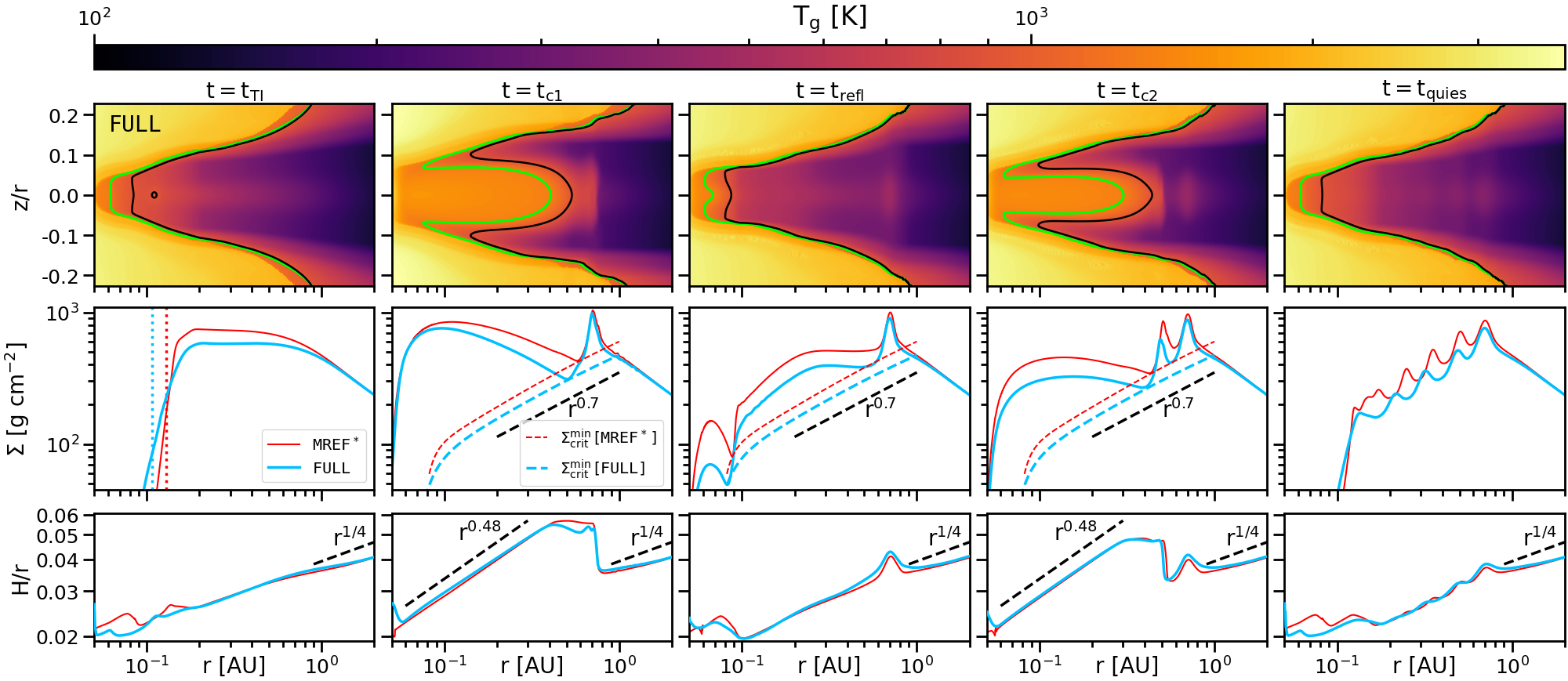}}
    \caption{Evolution of the outburst in the \texttt{FULL} model compared to $\texttt{MREF}^*$. The different columns correspond to different evolutionary stages, starting from the ignition of the burst at $t=t_\mathrm{TI}$, chronologically proceeding through the burst stage and ending with the beginning of the next quiescent state at $t=t_\mathrm{quies}$. The top row shows the temperature maps of \texttt{FULL} with the black and green contour lines marking the MRI-transition and the dust sublimation front. The middle row depicts the surface densities for both $\texttt{MREF}^*$ and \texttt{FULL} at the same stages during their respective evolution. The vertical dotted lines in the first panel indicate the locations of the DZIE in the corresponding model of the same colour. The panels in the second, third and fourth column also show the respective profiles of $\Sigma_\mathrm{min}^\mathrm{crit}$ together with a reference power law profile of $r^{0.7}$. The bottom row represents the aspect ratios $H/r$ of both models. The $r^{1/4}$ profile indicates the slope of $H/r$ in the outer disc (>2 AU). In the high-state regions, an additional reference profile of $r^{0.48}$ is shown.}
    \label{fig:burststructure}
\end{figure*}

\begin{figure*}[h]
    \centering
         \resizebox{\hsize}{!}{\includegraphics{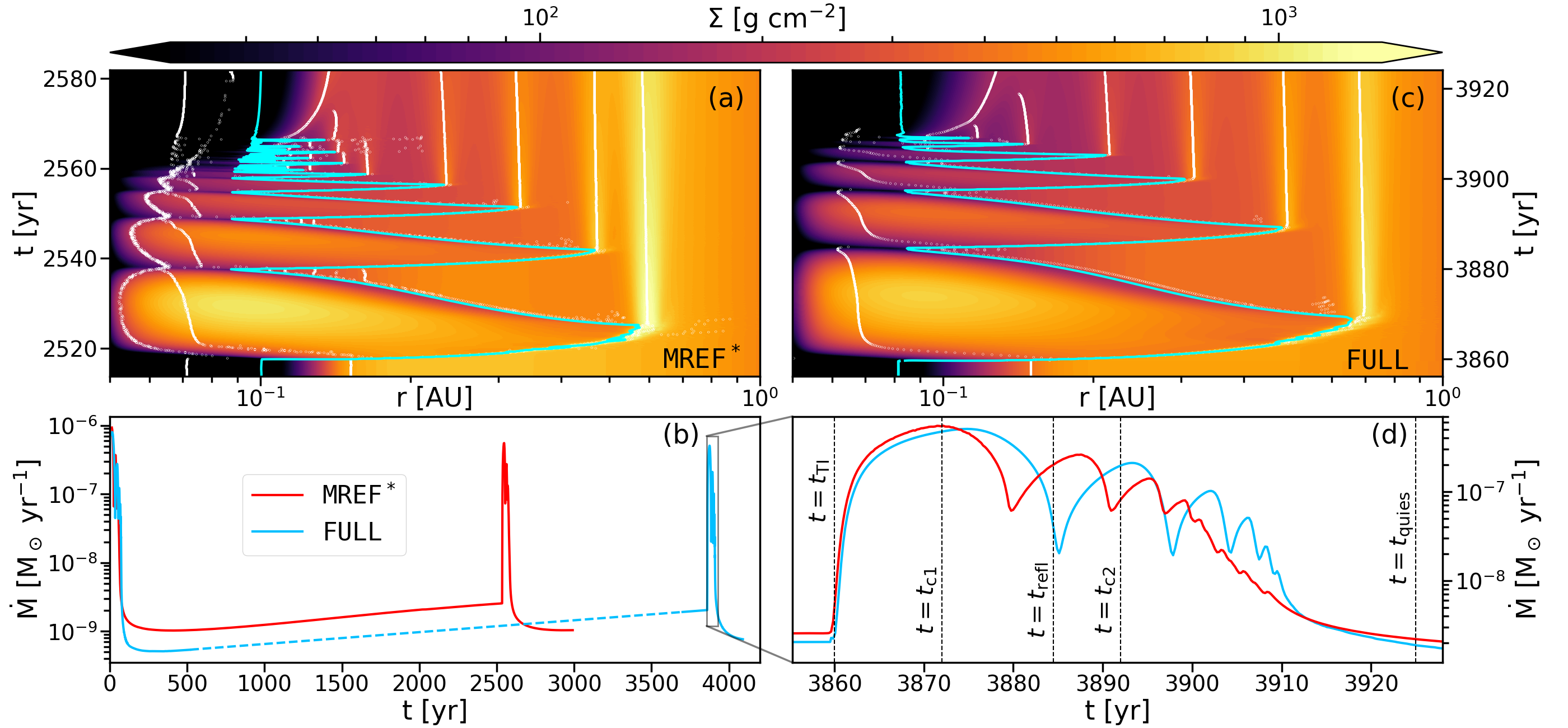}}
    \caption{Evolution of the surface densities and accretion rates. Panels (a) and (c) show the space-time diagrams of $\texttt{MREF}^*$ and \texttt{FULL}, respectively, where the displayed timespans have the same length. The white contours mark the positions of local pressure maxima, while the cyan line indicates the DZIE at the midplane. Panel (b) shows the accretion rates for both models over their entire simulation times. The dashed section of the blue line represents the interpolation between the two bursts. A magnification of the burst occurring in \texttt{FULL} is provided in panel (d). The vertical dashed lines mark the time of the ignition of the outburst ($t_\mathrm{TI}$), the first and second reflections of the heating front ($t_\mathrm{c1}, \, t_\mathrm{c2}$), the launching of the first reflare ($t_\mathrm{refl}$) and the beginning of the quiescent phase ($t_\mathrm{quies}$) for \texttt{FULL}. For comparison, the accretion rate resulting from the burst in $\texttt{MREF}^*$ is overplotted, shifted in time so that the respective $t_\mathrm{TI}$ for both models align.     }
    \label{fig:radius_time}
\end{figure*}
\subsection{Temporal evolution of the outburst phase} \label{sec:res_burst}

As a next step in analysing the influence of the new opacity descriptions, we compare the structure and evolution of the MRI-activated TI burst cycle following the quiescent phase between the models \texttt{FULL} and $\texttt{MREF}^*$.
Fig. \ref{fig:burststructure} shows five snapshots in time illustrating the evolution of a complete outburst cycle. The top row presents the two-dimensional temperature structure of \texttt{FULL}, analogous to that shown for \texttt{MREF} in \citetalias{Cecil2024b}, but on a logarithmic temperature scale. At $t = t_\mathrm{TI}$, the MRI is activated in the dead zone at the midplane, just outside the DZIE. The sudden increase in heating efficiency by viscous dissipation launches heating fronts into the dead zone, progressively activating the MRI. The front stalls when the surface density valley trailing it reaches $\Sigma_\mathrm{crit}^\mathrm{min}$ (at $t = t_\mathrm{c1}$), below which the MRI can no longer be sustained. It is then reflected into a cooling front that travels back towards the star, shutting down the MRI activity. This process repeats several times as smaller reflares (the first of which is ignited at $t=t_\mathrm{refl}$ and expands until $t=t_\mathrm{c2}$) until enough material has drained from the inner disc onto the star to prevent further MRI activation in the dead zone. The quiescent phase is then re-established at $t = t_\mathrm{quies}$.\par
The second row of Fig.~\ref{fig:burststructure} depicts the corresponding surface densities and radial profiles of $\Sigma_\mathrm{crit}^\mathrm{min}$ for both the \texttt{FULL} and $\texttt{MREF}^*$ models. The first panel shows that at $t=t_\mathrm{TI}$, \texttt{FULL} requires less overall mass in the inner disc for the conditions of MRI activation to be fulfilled. The reason for this difference is twofold: On the one hand, the DZIE and, consequently, the location of the first MRI activation in the dead zone reside at smaller radii, where the viscous heat dissipation is more efficient (despite the different choice of $T_\Delta$, the effects of which are explored in Appendix \ref{app:alphathings}). On the other hand, the larger dust Rosseland mean opacities in the dead zone increase the effectiveness of heat trapping near the midplane, which also shifts the profile of $\Sigma_\mathrm{crit}^\mathrm{min}$ to lower values. However, the slope of $\Sigma_\mathrm{crit}^\mathrm{min}$ remains mostly unchanged. Remarkably, the combination of a less massive inner disc at $t=t_\mathrm{TI}$ and the shift of $\Sigma_\mathrm{crit}^\mathrm{min}$ results in the outermost density- and pressure bump produced by the TI cycle being placed at the same location ($r\sim 0.7 \; \mathrm{AU}$) in both models. \par
At $t=t_\mathrm{refl}$, the density between the retreating cooling front and the star has to be much smaller in the \texttt{FULL} model in order for the heating by irradiation to become dominant and stop the retreat of the cooling front. This inability for further cooling closer to the star ultimately leads to the reignition of the MRI behind the stalled cooling front and the development of the reflares. After each reflare has placed a density maximum within the inner dead zone, the surface density retains an overall increasing slope in the regions affected by the burst at the beginning of the quiescent phase.  \par
The third row of Fig.~\ref{fig:burststructure} reveals that the aspect ratio is very similar in both models during the evolution of the outburst. In the regions where the equilibrium on the upper branch of the S-curve is established (see Sect.~\ref{sec:dis_Scurve}), $H/r$ adopts a radial power law with an exponent of 0.48, as already recognised in \citetalias{Cecil2024b}. This relation is a consequence of the temperature on the upper branch being roughly constant with radius (with a small dependence on surface density, as discernible in the S-curve shown and analysed in Sect. \ref{sec:dis_Scurve}). Following the definition of $H$, $H/r\propto T^{1/2}r^{3/2}r^{-1}\propto r^{1/2}$, which is consistent with our finding. With this scaling, the aspect ratio at a radius of $0.5\,\mathrm{AU}$ can be increased by a factor of up to two during the burst phase. \par
A more detailed evolution of the surface densities, as well as the accretion rates of the models \texttt{FULL} and $\texttt{MREF}^*$, is illustrated in Fig.~\ref{fig:radius_time}. Comparing the space-time diagrams in panels (a) and (c) shows that while the individual flares occupy more time in the case of \texttt{FULL}, the duration of the entire burst cycle is approximately equal in both models. As investigated in \citetalias{Cecil2024b}, the pressure bump at the DZIE in the quiescent phase is destroyed by the TI cycle and re-established after the burst has ended. During the accretion event, multiple pressure bumps are placed throughout the inner disc by the individual flares. The new opacity description does not have any significant effect on the placement and evolution of these maxima.\par
Panel (b) of Fig.~\ref{fig:radius_time} displays the evolution of the accretion rates of both models over their entire simulation time, which includes the initial TI cycle, the quiescent phase in which the inner disc is refilled by accretion and the viscous evolution of the dead zone, and the subsequent outburst. For the \texttt{FULL} model, the simulation of the quiescent phase had to be partly interpolated between the burst cycles due to computation time restrictions. We lay out the details as well as the consequences for the timescales of the quiescent phases in Appendix \ref{app:timescales}. It is expected, however, that the refilling of the material at the DZIE is slower in the \texttt{FULL} model due to the smaller extent and lower temperature of the MRI active region at the midplane in the quiescent phase compared to $\texttt{MREF}^*$ (and, equivalently, \texttt{DUST}, as shown in panel a of Fig.~\ref{fig:old_new_diff}). As a consequence, the total amount of angular momentum being transported outwards and taken over by the material behind the DZIE is decreased as well, making the accumulation of mass in the inner dead zone less efficient and prolonging the quiescent phase. \par
Panel (d) provides a more resolved view of the accretion rates during the episodic accretion events occurring in both models. The times at which the snapshots shown in Fig.~\ref{fig:burststructure} have been extracted are marked as vertical dotted lines for the example of \texttt{FULL}. As was already discernible in panels (a) and (c), the duration of the individual flares is increased in \texttt{FULL} due to the necessity of more material having to be accreted onto the star in the case of an optically thicker disc. The larger diminishment of density between the cooling front and the star, explained above in the context of Fig.~\ref{fig:burststructure}, manifests in panel (d) as the more pronounced dips in accretion rate between the reflares. Although the evolution of the accretion rates during the burst cycles differs between the two models, the maximum accretion rate is approximately the same, as is the total mass accreted onto the star with a value of $\sim\hspace{-0.3em}10^{-5}\,\mathrm{M_\odot}$.

\subsection{Influence of frequency-dependent irradiation and manifestation of the equilibrium temperature degeneracy} \label{sec:res_freqdep}
\begin{figure*}[h]
    \centering
         \resizebox{\hsize}{!}{\includegraphics{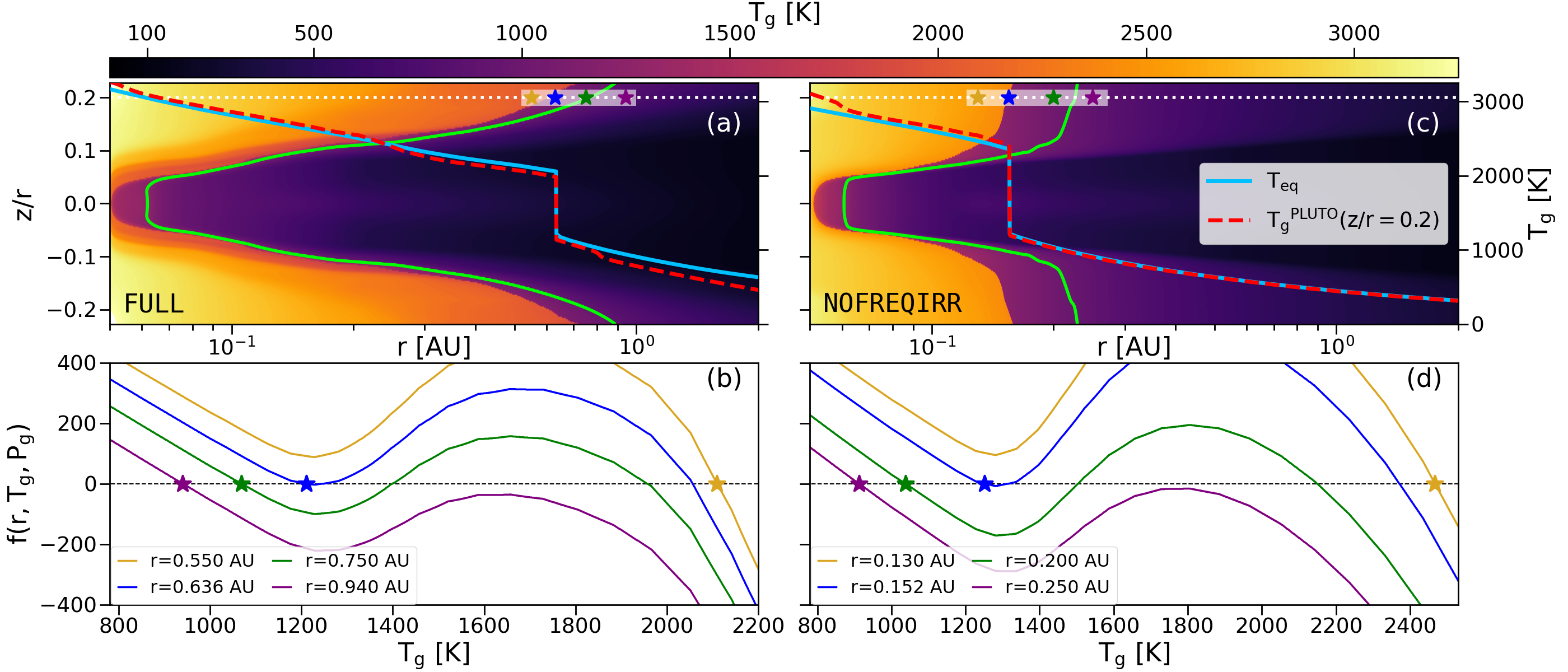}}
    \caption{Visualisation of the temperature jump as a consequence of the equilibrium temperature degeneracy. Panels (a) and (c) display temperature maps of the quiescent phase for the models \texttt{FULL} and \texttt{NOFREQIRR}, respectively. The red dashed lines show the radial profiles of the temperatures at a height of $z/r=0.2$ (indicated by the white dotted line). The solution of Eq. \ref{eq:T_eq} for $T_\mathrm{eq}$ at every radius is represented by the blue line. The green contours mark the dust sublimation fronts. Panels (b) and (d) show the function given in Eq. \ref{eq:f_Teq}, evaluated at four different radii for both models, respectively. The star-shaped markers indicate which of the available solutions is adopted by the solver at each radius. The same markers have been used to locate the respective evaluation radii in panels (a) and (c).}
    \label{fig:rootfinder}
\end{figure*}
Fig. \ref{fig:old_new_diff} and the top row of Fig.~\ref{fig:burststructure} include a remarkable detail inherent to the \texttt{FULL} model: a sudden drop in temperature, especially noticeable in the disc's atmosphere near the polar boundaries, that does not necessarily coincide with either of the transitions marked in the figures. Intuitively, it stands to reason that this feature might be a consequence of either the effects of frequency-dependent irradiation or the transition to an opacity dominated by dust. In order to show that this temperature drop is an inherent property of the gas Planck mean opacities, we compare the quiescent structure of \texttt{FULL} to the \texttt{NOFREQIRR} model in which we only consider the Planck mean opacities for the absorption of the irradiation flux. We aim to confront the temperature structures resulting from our numerical calculations with the expected equilibrium temperature $T_\mathrm{eq}$ of the optically thin medium in both models. A derivation of the analytic expression for $T_\mathrm{eq}$ is presented in Appendix \ref{app:Teq}. \par
The effect of frequency-dependent irradiation is illustrated in panels (a) and (c) of Fig.~\ref{fig:rootfinder}. While the midplane of the inner disc is largely unaffected, the two-dimensional temperature structure is altered significantly. In the \texttt{FULL} model shown in panel (a), the hot, gaseous atmosphere reaches higher temperatures close to the star and extends to much larger radii when compared to the temperature map of \texttt{NOFREQIRR} displayed in panel (c). However, the temperature drop is still manifested in \texttt{NOFREQIRR}, indicating already that its presence is not a consequence of frequency-dependent irradiation. The shape of the dust sublimation front (green contour) is affected as well by the altered position of the temperature jump. Additionally, the vertical temperature profile in the cool regions is more smoothed-out in \texttt{FULL}, in contrast to the very noticeable transition from the optically thick disc to the warmer atmosphere in \texttt{NOFREQIRR}. \par
Fig. \ref{fig:rootfinder} also analyses the origin of the temperature jump, which effectively separates the inner disk into a low and high temperature regimes. To efficiently calculate the analytic equilibrium temperature for comparison to the numerical values, we define the function $f(r,T_\mathrm{g},P_\mathrm{g})$ on the basis of Eq.~\ref{eq:T_eq},
\begin{equation} \label{eq:f_Teq}
    f(r,T_\mathrm{g},P_\mathrm{g})=\left( \frac{\kappa_\mathrm{irr}(T_\mathrm{g}, P_\mathrm{g})}{\kappa_\mathrm{P}(T_\mathrm{g}, P_\mathrm{g})}\right ) ^{-1/4}\left(\frac{R_\star}{2r} \right )^{1/2}T_\star - T_\mathrm{g} \; ,
\end{equation}
\noindent The roots of $f(r,T_\mathrm{g},P_\mathrm{g})$ represent the solutions for $T_\mathrm{eq}$. Panels (a) and (c) of Fig.~\ref{fig:rootfinder} compare the radial profiles of the temperatures of the underlying models at a height of $z/r=0.2$ (to ensure optical thinness) with the analytic equilibrium solution. We applied a root finding algorithm of the secant method to solve $f(r,T_\mathrm{g},P_\mathrm{g})=0$ at every radius. In order to isolate the consequences of the gas mean opacities, we excluded the influence of dust ($f_\mathrm{D2G}=0$), the varying profile of the gas pressure ($P_\mathrm{g}=\mathrm{const.}$) and the attenuation of the irradiation ($\tau_\mathrm{rad}(\nu_i,r)=0$) in the calculation of $T_\mathrm{eq}$. For the constant $P_\mathrm{g}$, we chose the mean value at $z/r=0.2$ of $6\cdot 10^{-7} \,\mathrm{dyn\,cm^{-2}}$. A full evaluation of $T_\mathrm{eq}$ with all contributions included in Eq.~\ref{eq:T_eq} is demonstrated in Appendix \ref{app:underrelax}.\par
The temperature profiles of our numerical models fit the analytic solution reasonably well in both the \texttt{FULL} and the \texttt{NOFREQIRR} models, considering the neglected effects described above in the analytic calculations. Most notably, the presence and location of the temperature jump can be exactly recreated in both models by the simplified analytic considerations. For further analysis, panels (b) and (d) show $f(r,T_\mathrm{g},P_\mathrm{g})$ as a function of $T_\mathrm{g}$ for four different radii, respectively, calculated with the same simplifications as indicated above. The gold lines depict $f(r,T_\mathrm{g},P_\mathrm{g})$ at a radius just before the location of the temperature jump. In this case, the only available solution for the equilibrium temperature is in the high-temperature regime. The blue curves were calculated at a larger radius where a second root becomes available, which is the solution adopted by our numerical solver algorithm. Starting from the blue profile, $f(r,T_\mathrm{g},P_\mathrm{g})$ allows for three solutions until the radius of the purple curve, after which only one solution for $T_\mathrm{eq}$ is allowed again. Consequently, for radii between the blue and the purple curves, the solution for the equilibrium temperature is degenerate. This phenomenon is equivalent to the equilibrium temperature degeneracy described by \cite{Malygin2014}. \par
The star markers indicate that our models immediately switch to the low-temperature solution as soon as it becomes available and do not revert to the high-temperature regime at larger radii. Arguably, this is a consequence of our chosen method of setting up the initial model. Since we start from a cool disc and let the temperature increase by radiative transport, the low-temperature solutions are conserved. Conversely, if we chose to start from a hot disc and let it cool down radiatively towards the initial model, the temperature jump would occur just before reaching the radius of the purple curves. As argued by \cite{Malygin2014}, the available intermediate solutions are always unstable and will therefore not be adopted by our solver algorithm. This is the reason for the visible gap in panels (b) and (d) of Fig.~\ref{fig:mean_opacs} in the coverage of the temperature-pressure pairs occurring in our simulations. Regardless of the choice of the initial model creation, the temperature jump intrinsically has to manifest at a radial range bounded by the blue and purple curves shown in panels (b) and (c) of Fig.~\ref{fig:rootfinder} in all models that include the gas mean opacities of \cite{Malygin2014}.

\begin{figure*}[h!]
    \centering
         \resizebox{\hsize}{!}{\includegraphics{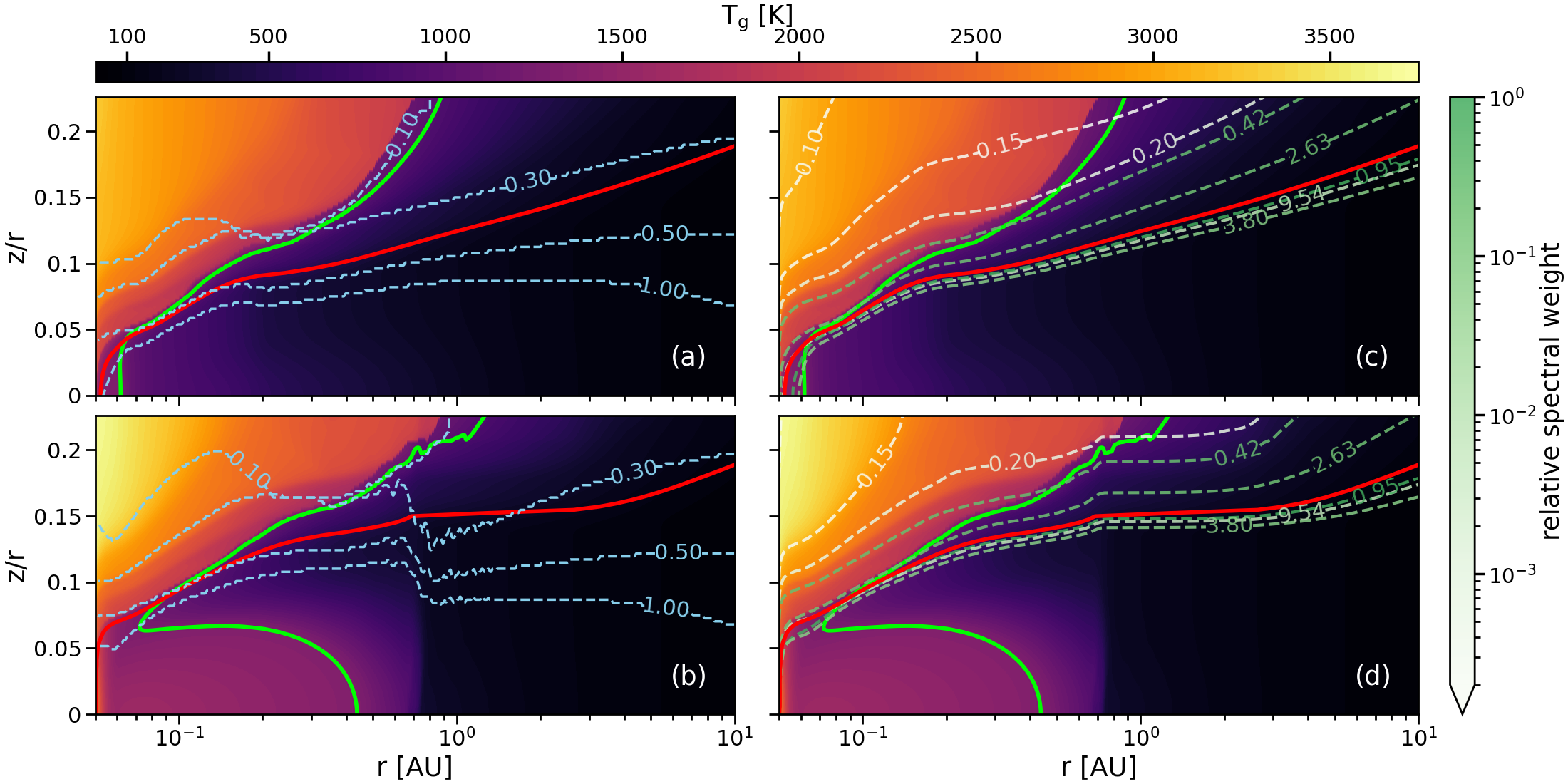}}
    \caption{Different $\tau=1$ lines for a disc in quiescence (panels a and c) and in outburst (panels b and d) shown atop the underlying temperature structure for the model \texttt{FULL}. In panels (a) and (b), the dashed blue lines mark the $\tau_\mathrm{vert}(\nu)=1$ surfaces for different wavelengths, which are indicated in the labels in units of $\mu$m. The dashed lines in panels (c) and (d) show the locations of the $\tau_\mathrm{rad}(\nu)=1$ transitions for the stellar irradiation for several representative wavelength bins with the respective central wavelengths in $\mu$m marked in the labels. The colour of the dashed lines indicates the bin's relative spectral weight. Additionally, each panel includes the dust sublimation front as the green line and the $\tau_\mathrm{rad,tot}=1$ line for the total radial optical depth in red. }
    \label{fig:tau1s}
\end{figure*}

\subsection{Locations of radial and vertical $\tau=1$ surfaces} \label{sec:res_tau1}
The availability of frequency-dependent opacities for both gas and dust allows us to calculate the optical depths for different wavelengths. Fig.~\ref{fig:tau1s} presents the profiles of the radial ($\tau_\mathrm{rad}(\nu)$, panels c and d) and vertical ($\tau_\mathrm{vert}(\nu)$, panels a and b) $\tau=1$ surfaces for selected wavelengths in the upper hemisphere of the \texttt{FULL} model, including accretion shock luminosity feedback. Both the quiescent (panels a and c) and outburst phases (panels b and d) are represented. \par
For the $\tau_\mathrm{vert}(\nu)=1$ surfaces, the optical depth was integrated along the (Cartesian) $z$-direction, starting from the upper polar boundary down towards the midplane. The positions of the surfaces for the different wavelengths can be understood by comparing them to the wavelength-dependent opacities shown in Fig.~\ref{fig:B_star}. At smaller wavelengths, radiation is more readily absorbed, especially due to the higher gas opacities. Remarkably, the disc can become optically thick even without the contribution of dust at the shortest wavelengths. Starting from about $1\,\mu\mathrm{m}$ and going to longer wavelengths, the surfaces lie deep within the disc at approximately the same position (for clarity, only $\tau_\mathrm{vert}(1\,\mu\mathrm{m})=1$ is shown). During a burst (panel b), the MRI active region inflates the disc significantly, pushing all $\tau_\mathrm{vert}(\nu)=1$ surfaces to greater heights. The effect of the accretion luminosity feedback is clearly noticeable as well by the higher temperatures in the atmosphere close to the star. However, its influence on both the dynamics and the position of the $\tau_\mathrm{vert}(\nu)=1$ lines remains mostly marginal. A small effect of the larger irradiation flux is the disturbance of the temperature jump caused by the equilibrium temperature degeneracy analysed in Sect.~\ref{sec:res_freqdep}. The effective increase in stellar temperature slightly shifts the temperature jump to larger radii. After the burst phase, these new equilibrium solutions remain conserved (as also evident in comparing the first and last panels in the first row of Fig.~\ref{fig:burststructure}). \par
In panels (c) and (d), the $\tau_\mathrm{rad}(\nu)=1$ have been calculated for the central wavelengths of a number of selected frequency bins. Generally, radiation in wavelength bins with high spectral weight can penetrate deeper into the disc, while radiation at small wavelengths is already mostly absorbed by the gas before reaching the dust sublimation front. As a consequence, the transition between optically thin and thick, considering the total stellar irradiation flux, $\tau_\mathrm{rad,tot}=1$, lies deeper within the disc as well. Similar to the case for the $\tau_\mathrm{vert}(\nu)=1$ lines, the $\tau_\mathrm{rad}(\nu)=1$ surfaces are shifted away from the midplane during the outburst phase. Consequently, the surfaces become horizontal in the region beyond the MRI active zone up to several AU, indicating the shadowing effect of the puffed-up, hot inner disc.

\subsection{Stability of burst features} \label{sec:res_RWI}
\begin{figure}[t!]
    \centering
         \resizebox{\hsize}{!}{\includegraphics{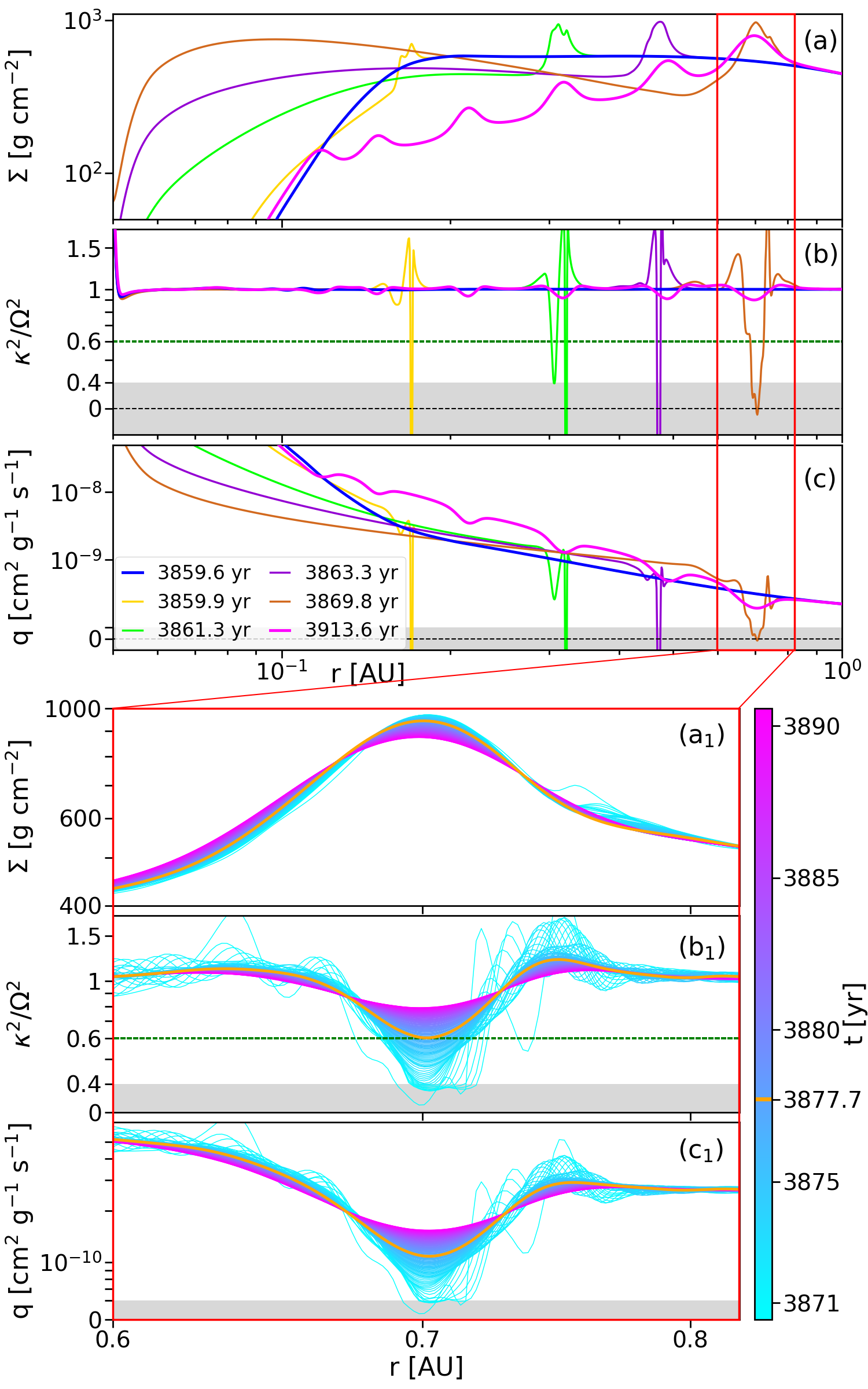}}
    \caption{Analysis of the stability criteria for the density maxima generated during the outburst in the \texttt{FULL} model. The top three panels display the temporal evolution of the surface density $\Sigma$ (panel a), the parameter $\kappa^2/\Omega^2$ (panel b) and the vortensity $q$ (panel c). These are shown as snapshots during the timeframe starting from the ignition of the burst ($t=t_\mathrm{TI}$, blue line) until the first density bump has reached its outermost position and $\kappa^2/\Omega^2$ becomes positive (brown line). Additionally, the pink line represents the disc at $t=t_\mathrm{quies}$. The green dashed line in panels (b) and ($\mathrm{b}_1$) indicates the halfway-to-Rayleigh stability threshold. The bottom three panels show the same three quantities as above, but zoomed into the radial region where the first density bump is located. The colour-coded timeframe begins at the moment when $\mathrm{min}(\kappa^2/\Omega^2)>0$ and includes the transition from instability to stability according to the halfway-to-Rayleigh criterion, reached at $t=3877.7$ yr (orange line in panels $\mathrm{a}_1$, $\mathrm{b}_1$ and $\mathrm{c}_1$ and in the colour bar). Quantities in panels (b), ($\mathrm{b}_1$), (c) and ($\mathrm{c}_1$) are depicted on a logarithmic scale with a linear transition around zero, indicated by the grey shaded region. }
    \label{fig:RWI}
\end{figure}

The emergence of steep density features during a burst cycle (e.g. middle row of Fig.~\ref{fig:burststructure}) raises questions about their dynamic stability.
We emphasise that the RWI can not be manifested in our axisymmetric models. For the analysis of the stability of the DZIE and the burst features, we rely on criteria that assess the linear RWI. A complete investigation of this aspect of the inner disc's evolution, including non-linear effects, would require non-axisymmetric simulations, ideally in 3D. 
Fig.~\ref{fig:RWI} shows an analysis of the density bumps developing during the burst in the \texttt{FULL} model with respect to potential Rossby wave and Rayleigh instability. In this context, we consider the Lovelace criterion \citep{Lovelace1999, Lovelace2014} as a necessary condition for RWI, the halfway-to-Rayleigh limit \citep{Chang2024} as a sufficient criterion and the condition resulting from the fit to the amplitude of marginally stable features according to \cite{Ono2016} for comparison.  \par
For the Lovelace criterion to be fulfilled, extrema in the vortensity profile $q$ of the disc have to be present, while the halfway-to-Rayleigh condition is given as $\mathrm{min}(\kappa_\mathrm{ef}^2)/\Omega^2\lessapprox0.5$-$0.6$ with $\kappa_\mathrm{ef}$ being the epicyclic frequency. In order to compare the density bumps occurring in our models with the descriptions of marginally stable features given in Table 2 of \cite{Ono2016}, we approximate them by Gaussian bumps with corresponding amplitudes and widths. For the definitions of the relevant quantities for the three considered criteria, we refer to Appendix \ref{app:RWI_expressions}.\par
The first expansion of the heating front during the accretion event occurring in the \texttt{FULL} model is analysed in the top three panels of Fig. \ref{fig:RWI}. At $t=t_\mathrm{TI}$, no extrema in the vortensity profile are present. Even the DZIE does not fulfil any of the three criteria. If the density contrast at the DZIE increased further, the RWI might still become active. However, this is prevented by the emergence of the accretion event. The density peak travelling outwards alongside the heating front produces sharp minima in both the vortensity and $\kappa_\mathrm{ef}^2/\Omega^2$ profiles, fulfilling both the Lovalace and halfway-to-Rayleigh criteria. Since $\kappa_\mathrm{ef}^2/\Omega^2$ becomes negative, these features can even be considered to be Rayleigh unstable. After the heating front has stalled, Rayleigh stability is quickly restored and the density feature starts to diffuse on the viscous timescale. A part of this process is shown in the bottom three panels of Fig. \ref{fig:RWI}. The time between the density feature reaching its outermost position and the transition to stability according to the upper limit of the halfway-to-Rayleigh criterion (orange line) is around seven years, which corresponds to $\sim \hspace{-0.14cm}12$ local orbits at the central radius of the bump. During this time period, the amplitude of the feature is large enough to also render it unstable according to the criterion proposed by \cite{Ono2016}. Regarding this condition, the RWI remains even after the halfway-to-Rayleigh threshold is crossed. However, the width of the density bump is always larger than the local scale height, which should make the halfway-to-Rayleigh criterion more accurately applicable in this timeframe \citep{Chang2024}.\par
By the time the burst cycle has completed (at $t=t_\mathrm{quies}$), the vortensity profile still shows minima at the locations of the density bumps. However, the inner disc is stable to the RWI throughout its entire extent according to both the halfway-to-Rayleigh and \cite{Ono2016} criteria.\par
Although these conditions should not directly be influenced by the levels of turbulence \citep{Lin2013}, the steepness of the density structure at the DZIE and the evolutionary timescale of the features placed within the dead zone are subject to the description of the viscous $\alpha$ parameter. An investigation of the effect of a smaller value of $\alpha_\mathrm{DZ}$ and a sharper transition at the DZIE on the RWI of the inner disc is presented in Appendix \ref{app:alphathings}.

\section{Discussion}\label{sec:discussion}
The models analysed in this work show that although the thermal structure of the inner disc is crucially affected by the careful treatment of dust and gas opacities, the periodic instability mechanism disrupting the inner disc and leading to accretion outbursts is operational in all cases, and hence denotes a robust feature of disc evolution. In this section, we discuss the potential observational prospects expected from our simulations and address the possible consequences of the RWI and Rayleigh instability of burst features. We further analyse the dominance of the dust opacities during the burst evolution, indicated by our results, by investigating the S-curves of different models. Finally, we assess the importance of considering detailed dust and gas opacities for our understanding of the inner disc structure and its evolution before describing the limitations of our models.
\subsection{Preview of potential observational consequences}
The significant changes in the temperature structure of the inner disc resulting from our treatment of the gas opacities may have considerable implications for the observational signatures produced. While a more detailed analysis will be part of future work, we aim to provide a first estimate of the expected observational impacts in this section. \par
Concerning the spectral energy distribution (SED), we may expect a larger flux in the near infrared coming from the hot gas atmosphere, filling in the spectral region where dust does not contribute due to sublimation. In the mid to far infrared, changes in the SED may be insignificant, as these regions are primarily dominated by dust emission. The implementation of temperature- and density-dependent gas opacities in radiative transfer codes, such as \texttt{RADMC-3D} \citep{Dullemond2012}, is non-trivial and warrants a dedicated study. To still assess the influence of the altered inner disk structure by the new opacity descriptions on the resulting SEDs, we adopted a simplified approach in the following. \par
Instead of determining the absolute values of the SEDs, we only intend to estimate the differences in the emissions of the $\texttt{MREF}^*$ and \texttt{FULL} models by calculating,
\begin{equation} \label{eq:SEDdiff}
    \overset{\texttt{FULL}}{\underset{\texttt{MREF}^*}{\Delta}}F_\nu=\frac{{F}_\nu(\texttt{FULL})-F_\nu(\texttt{MREF}^*)}{F_\nu(\texttt{MREF}^*)} \;,
\end{equation}
where the frequency-dependent flux is evaluated for each model with \citep{Chiang1997},

\begin{equation}
    F_\nu=k\;\int_{r_\mathrm{in}}^{r_\mathrm{out}} r  \int_{-\infty}^{\infty} \frac{\partial \tau_\mathrm{vert}(\nu,r, z)}{\partial z}e^{-\tau_\mathrm{vert}(\nu,r,z)}B_\nu(\nu,T_\mathrm{g})\,dz\,dr \;,
\end{equation}
with $z$ being the cylindrical height and $k$ a constant factor that is cancelled when evaluating Eq. \ref{eq:SEDdiff}. For the calculation of $\tau_\mathrm{vert}(\nu,r, z)$, we took the contributions of both the dust and gas opacities into account. This method assumes that both the gas and dust components are radiating as blackbodies at their respective local temperatures. 
\par
\begin{figure}[t]
    \centering
         \resizebox{\hsize}{!}{\includegraphics{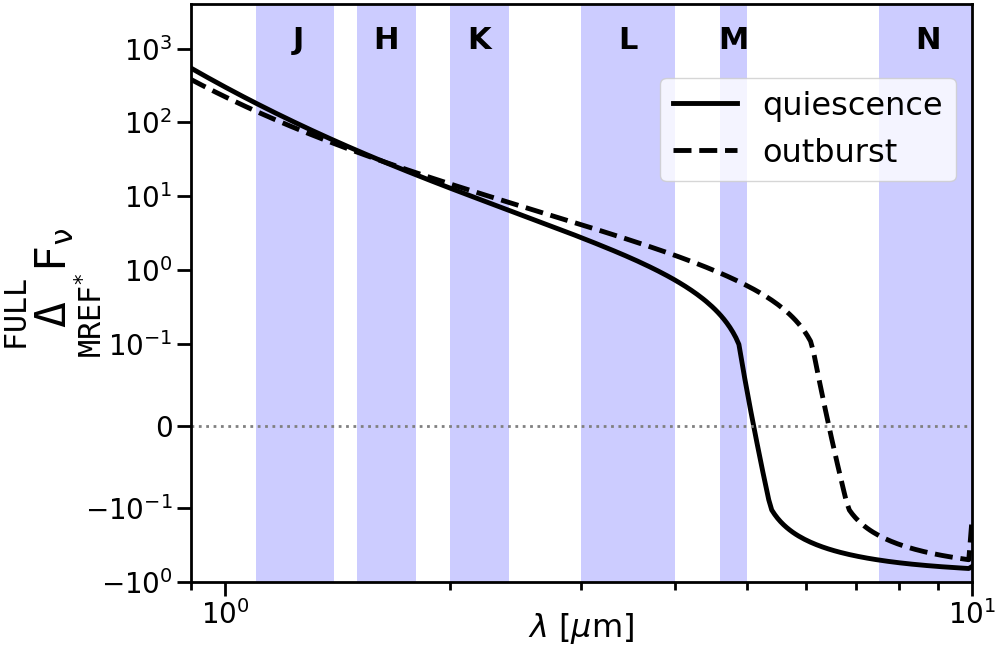}}
    \caption{Differences in the SEDs between the $\texttt{MREF}^*$ and \texttt{FULL} models during both quiescent and outburst phases. The shaded areas mark the wavelength regions covered by the various infrared bands. }
    \label{fig:SED}
\end{figure}
Fig. \ref{fig:SED} shows the results of Eq. \ref{eq:SEDdiff} for both the quiescent and outburst phases in the near to mid infrared spectral region, with markings for the main infrared bands. The flux in the near infrared is significantly enhanced in the \texttt{FULL} model compared to \texttt{MREF} by virtue of the hotter disc atmosphere. Starting from the M-band, the flux in \texttt{MREF} is slightly increasing above the flux emerging from \texttt{FULL} due to the less massive inner disk in \texttt{FULL} and the rise in influence of the dust opacities at larger wavelengths. The differences between the quiescent and outburst phases are only marginal, with the dominance of the \texttt{FULL} model shifting slightly towards the mid infrared region. This simplified approach presents a first idea of the significance of the change in the structure of the inner disc by considering (primarily) the detailed gas opacities on the spectral emissions. However, to determine the precise shape of the SEDs resulting from our models, dedicated radiative transfer calculations with careful implementation of the gas opacities and, ideally, thermochemical considerations in the hot disc corona are necessary.
\par
Apart from effects on the SEDs, the hot gas in the disc atmosphere could potentially lead to the emergence of several line emissions. For instance, CO overtone emissions have previously been associated with a hot gas component with temperatures of 3500\,K in the innermost regions of Class I discs \citep{Lee2016, Rudy2023}. This temperature regime can potentially be reached in our models in the thin disc atmosphere (e.g., Fig.~\ref{fig:tau1s}). \par
The disc regions close to the $\tau_\mathrm{rad,tot}=1$ surface with temperatures in the range of 1500\,K to 3000\,K could fulfil the conditions for producing $\mathrm{H_2}$ rovibrational line emission in the NIR by collisional excitation in the dense, hot gas \citep{Beck2019}. Furthermore, the hot, thin gas atmosphere may also enable forbidden line emissions, particularly of [$\mathrm{O}\,\mathrm{I}$], which, for instance, has been observed to originate from the inner regions within the disc of ET Cha \citep{Woitke2011}. However, forbidden line emissions are typically also expected to be produced in jets and winds emerging from the inner disc \citep[e.g., ][]{Flores-Rivera2023}. The true origin might be discernible by disentangling different velocity components of the lines. \par
The high-resolution spectrum of the frequency-dependent gas opacities calculated by \cite{Malygin2014} potentially allow us to track the strength and origin of specific line emissions in the near- and mid-infrared, as would be observable by instruments such as NIRSpec or MIRI of the James Webb Space Telescope. A detailed investigation of the implications for both the SEDs and the line emission spectra using tools for radiative transfer modelling, such as \texttt{RADMC-3D}, goes beyond the scope of this study and is left for future work.

\subsection{Possible implications of the instability of density features}
As shown in Sect. \ref{sec:res_RWI}, the density spikes travelling outwards alongside the expanding MRI active region are prone to the Rayleigh instability, while the bumps left behind in the inner disc continue to fulfil the conditions for RWI for more than ten local orbits. 
The RWI has been recognised as an effective production mechanism for vortices at the edges of planetary gaps \citep[e.g.,][]{Li2005, Hammer2019, Ziampras2025a} or at strong transitions in turbulence such as the edges of dead zones \citep[e.g.,][]{Faure2014, Flock2015, Roberts2025}. The duration of the linear growth phase of the RWI is on the order of ten local orbits, after which vortices are readily formed \citep{Lyra2012, Bae2015}. However, this mechanism is intrinsically non-axisymmetric and does not manifest in our models. Based on the analysed RWI criteria and the timescales of linear growth and saturation of the instability, we conclude that the density bumps placed by the outburst mechanism, including its reflares, will potentially lead to the emergence of vortices in a non-axisymmetric geometry. However, such vortices typically merge together quickly to form a single vortex \citep[e.g.,][]{Bae2015} and can decay on short timescales due to processes such as viscous diffusion (governed by $\alpha_\mathrm{DZ}$), slow cooling or dust dynamics \citep{Rometsch2021, Ziampras2025b, Ziampras2025a}. Vortices produced by RWI at the edges of dead zones can potentially survive longer \citep{Regly2017}, but the results of our work indicate that the DZIE seems to be mostly stable. Therefore, we expect the vortices resulting from RWI in the density bumps to decay on a short timescale while new vortices are continuously produced by the ongoing burst mechanism. A smaller value of $\alpha_\mathrm{DZ}$ could potentially allow the vortices to linger in the dead zone for a longer time. \par
Rayleigh instability of density features effectively results in their radial diffusion on a dynamic timescale \citep[e.g.,][]{Yang2010}. Since the formation of vortices would also effectively smooth out sharp density features, it is reasonable to assume that the bumps placed in the inner disc by the burst mechanism will not be as pronounced as our models predict (Fig. \ref{fig:burststructure}). \par
The question remains of whether the Rayleigh or Rossby wave unstable, outward-moving density spikes pose an impediment to the expansion of the heating front. The progression of the heating front relies in part on the activation of the MRI by increasing the density ahead of the front above a critical value. Consequently, if the timescale of the dispersion of the density spikes by the instabilities is comparable to the timescale of the expansion of the MRI active region, the heating front may be slowed down and could possibly be stalled sooner. However, neither the possible impediment of the heating front nor the quicker dispersion of the density bumps will significantly affect the emergence of the burst cycles and the resulting density structure within the quiescent phase. Non-axisymmetric simulations are needed to investigate the interplay between these different instability dynamics in more detail.

\subsection{Alteration of S-curves} \label{sec:dis_Scurve}
\begin{figure}[t]
    \centering
         \resizebox{\hsize}{!}{\includegraphics{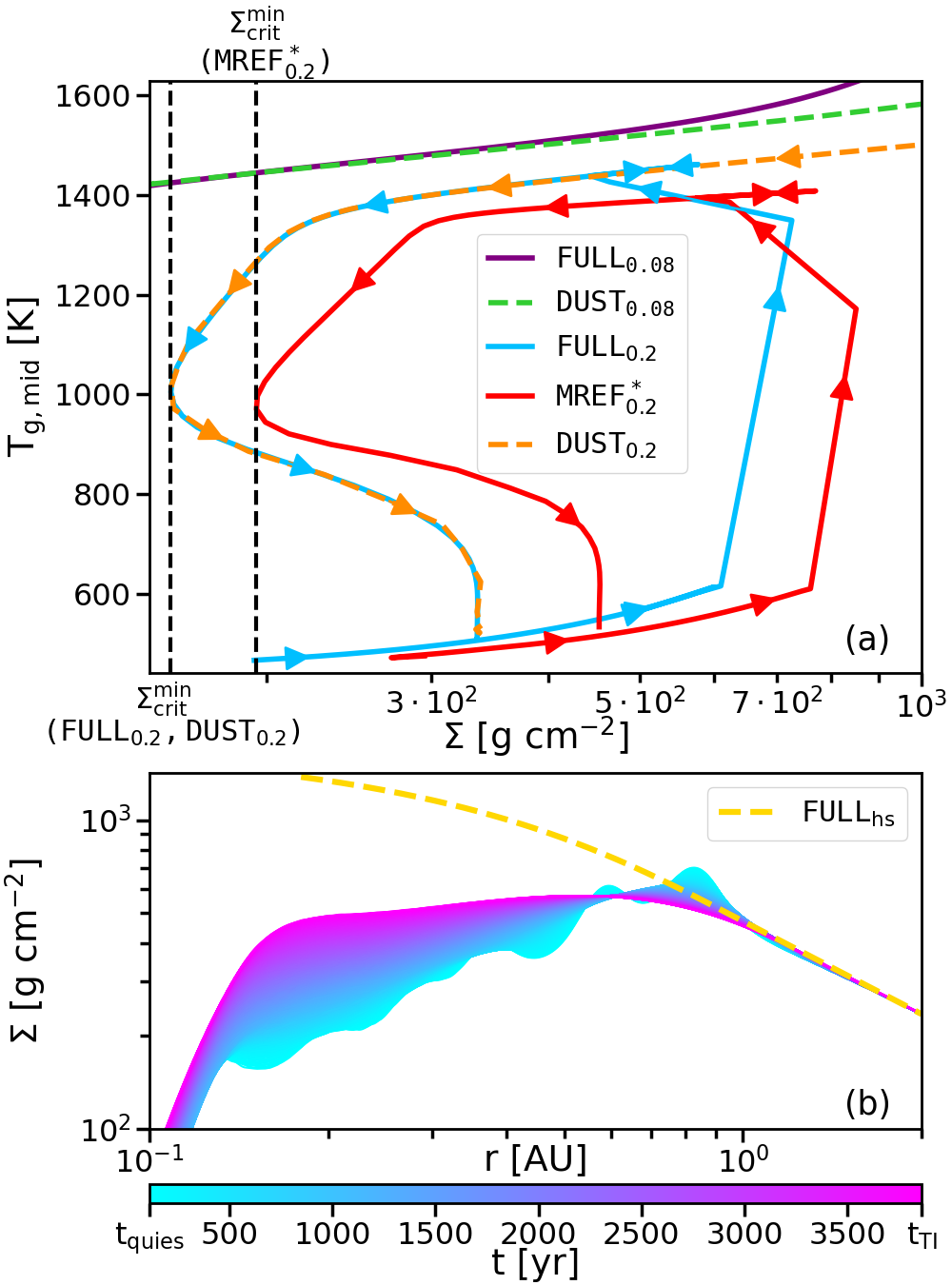}}
    \caption{S-curve behaviour with respect to the MRI-activated TI and evolution of the inner disc during quiescence. Panel (a) compares the S-curves of $\texttt{MREF}^*$, \texttt{DUST} and \texttt{FULL}, extracted from the respective quiescent phases (for $\texttt{MREF}^*$ and \texttt{FULL}) and the main flares of the accretion events (for all three models) at a radius of $0.2\,\mathrm{AU}$. The arrows along the curves indicate the direction of evolution. The black dashed lines mark the critical values of the surface densities at which the respective upper branches end. The panel also includes the upper branches of the S-curves for the \texttt{DUST} and \texttt{FULL} models at $r=0.08\,\mathrm{AU}$. The evolution of the surface density during the quiescent phase of \texttt{FULL} is displayed in panel (b). The yellow dashed line is the approximate profile of the hydrostatic structure the disc would adopt if no TI were to occur.  }
    \label{fig:Scurve}
\end{figure}

A major part of \citetalias{Cecil2024b} revolved around the tracing of the episodic accretion events along S-curves of thermal stability. Panel (a) of Fig. \ref{fig:Scurve} illustrates the influence of the new opacity descriptions on the shapes and positions of the curves in the $T_\mathrm{g,mid}$-$\Sigma$ plane. The S-curves are shown for the $\texttt{MREF}^*$, \texttt{DUST} and \texttt{FULL} models, calculated at a radius of $0.2\,\mathrm{AU}$. They include the first flare of the burst cycle\footnote{While the curve originates from the initial TI cycle for \texttt{DUST}, the $\texttt{MREF}^*$ and \texttt{FULL} curves result from the burst after the quiescent phase. This is why the upper branch of \texttt{DUST} extends to larger surface densities.}, omitting the reflares for clarity. \par
The discrepancies between the $\texttt{MREF}^*$ and \texttt{FULL} curve are another manifestation of what has already been recognised in Sects.~\ref{sec:res_dust}--\ref{sec:res_burst}. In \texttt{FULL}, the TI is initiated at smaller surface densities, the temperature in the MRI active region (high-state, upper branch of the S-curve) is higher and $\Sigma_\mathrm{crit}^\mathrm{min}$ lies at smaller values. As described in \citetalias{Cecil2024b} and also discussed in other studies \citep[e.g.][]{Woitke2024}, the upper branch of the S-curve is shaped by a thermostat effect. Thereby, a continuous rise in temperature is prevented by the sublimation of dust, allowing the increased radiative cooling to counteract the viscous heat dissipation and establishing the high-state equilibrium. The slopes of the upper branches only change at temperatures well below the dust sublimation threshold, where the dust-to-gas mass ratio is constant. \par
Curiously, the curves for \texttt{DUST} and \texttt{FULL} overlap exactly at the chosen radius. The underlying reason has already been mentioned in the context of Fig. \ref{fig:old_new_diff}. In the optically thick MRI active region, the radiative cooling is determined by the Rosseland mean opacities, which will always be dominated by the dust contribution as long as a sufficient amount of dust is still present (see Fig.~\ref{fig:mean_opacs}). Since this is the case at $0.2\,\mathrm{AU}$, the gas opacities do not influence the S-curve at this location. They only become relevant when the temperatures reach values high enough for the majority of the dust to be sublimated, such that the values of $\kappa_\mathrm{R,g}$ and $f_\mathrm{D2G}\kappa_\mathrm{R,d}$ become comparable. This effect is revealed when calculating the S-curves closer to the star, where heating becomes more efficient. Panel (a) of Fig. \ref{fig:Scurve} includes the upper branches of \texttt{DUST} and \texttt{FULL} at $r=0.08\;\mathrm{AU}$, showing that they indeed start to diverge at high temperatures. The same effect would manifest even at larger radii if the disc were chosen to be more massive. \par
Along the lower branches, the discs evolve slowly on the viscous timescale, $t_\nu=r^2/\nu$, determined by $\alpha_\mathrm{DZ}$. Panel (b) of Fig.~\ref{fig:Scurve} shows the evolution of the surface density during the quiescent phase of the \texttt{FULL} model. It illustrates the refilling of the inner disc by accretion of material from larger radii and the diffusion of the density bumps produced by the previous accretion event. If no further burst cycles were to occur, the refilling would proceed until a hydrostatic, steady-state solution is reached, an estimate of which is shown as the yellow dashed line. However, the conditions for TI are fulfilled at $t=t_\mathrm{TI}$ and the surface density structure is reset by the burst to approximately what is shown in panel (b) at $t=t_\mathrm{quies}$.

\subsection{The necessity of including detailed opacity descriptions} \label{sec:dis_opacnecessity}
Section~\ref{sec:results} provides comparisons of simulations with different opacity descriptions. The confrontation of these models with each other allows us to gauge the necessity of complex and computationally expensive opacity descriptions for different use cases. \par
Sections~\ref{sec:res_dust}--\ref{sec:res_burst} and \ref{sec:dis_Scurve} make clear that the evolution and extent of the episodic accretion event caused by MRI activation in the dead zone is only influenced by the mean dust opacity description, while the gas opacities and the frequency-dependence of the irradiation do not play a major role. However, the position of the DZIE and, consequently, the location of the ignition of the outburst and the efficiency of the density pile-up during quiescence are strongly affected by the gas contribution. Therefore, it cannot be assumed that the duration of the quiescent phase and the shape of the disc at $t=t_\mathrm{TI}$ are independent of the gas opacities. Furthermore, as indicated in Sect. \ref{sec:dis_Scurve}, for more massive discs and stronger viscous heating during the burst, the gas mean opacities could become the dominant contributor to the cooling efficiency if dust is sufficiently sublimated. Therefore, concerning the burst mechanism investigated in this work and \citetalias{Cecil2024b}, we propose that if the duration of the quiescent phase and the structure of the inner disc at the ignition of the TI are of interest, the gas mean opacities should not be neglected. For the burst phase itself, the careful treatment of just the mean dust opacities should suffice, but only as long as the thermostat effect on the upper branch of the S-curve can be sustained throughout the majority of the burst region. \par
The axisymmetric temperature structure of the inner disc is majorly impacted by the gas opacities. As displayed in Fig. \ref{fig:old_new_diff}, the temperature in the disc's atmosphere increases drastically when gas opacities are treated carefully. The frequency-dependent irradiation exacerbates this behaviour even further (Fig. \ref{fig:rootfinder}). The radial temperature profile in the very inner disc in the vicinity of the midplane is strongly affected as well. While in the simpler $\texttt{MREF}^*$ model, the dust sublimation front is located outside $0.1\,\mathrm{AU}$, it shifts to smaller radii (0.06--0.07\,AU) in the \texttt{FULL} model, which agrees better with recent estimates of inner dust wall locations in discs around classical T Tauri stars \citep{Pittman2022}. Furthermore, the gas mean opacities evoke the manifestation of the equilibrium temperature degeneracy as a sharp jump in temperature, as analysed in Fig. \ref{fig:rootfinder}. The location of the transition between high- and low-temperature regimes is significantly impacted by the consideration of frequency-dependent irradiation. Additionally, the determination of positions of different $\tau=1$ surfaces, using frequency-dependent opacities, makes it possible to gauge where observed radiation of different wavelengths originates from. Therefore, for investigating the two-dimensional temperature structure of the inner disc and analysing its observational consequences, gas opacities with both their mean and frequency-dependent values are of significant importance. \par
However, it is worth pointing out that including detailed opacity descriptions in simulations can come at considerable computational costs. Considering that the gas opacities depend on temperature and pressure in both their mean and frequency-dependent versions, the necessary interpolations in each cell for every frequency bin can severely increase the computational load. For instance, all aspects of the computational setup being equal, the computational time for the \texttt{FULL} model was larger by a factor of around 2.4 in comparison to $\texttt{MREF}^*$.

\subsection{Model limitations} \label{sec:dis_limits}
A major part of the limitations of the models of \citetalias{Cecil2024b} has been addressed in this work. However, several physical aspects, which may have an influence on our results, were not taken into account. \par
In the equation of state (Eq. \ref{eq:state}), we assumed a constant value for $\mu_\mathrm{g}$ of 2.35 for a mixture of hydrogen and helium with solar abundance, following \citet{Flock2019}. The same assumption was used for fixing the first adiabatic index $\gamma$. However, $\mu_\mathrm{g}$ and $\gamma$ typically depend on the ionisation state of the material in addition to the chemical composition \citep[e.g.][]{DAngelo2013}. While our assumption may still be reasonable for models with maximum temperatures around the dust sublimation limit (as in $\texttt{MREF}^*$), it breaks down in the hot gaseous regions occurring in simulations with detailed gas opacities. In our models, the absorbed energy in the thin atmosphere translates directly into an increase in temperature, rather than being also invested in the partial ionisation of the gas at high temperatures. Additionally, energy may be expended for the endothermic process of $\mathrm{H}_2$ dissociation at temperatures larger than 2000\;K. These simplifications could result in an overestimation of the temperatures in the atmospheres of the \texttt{FULL} and \texttt{NOFREQIRR} models. However, the ionisation state and its influence on the equation of state have been taken into account in the calculations of the gas opacities by \cite{Malygin2014}. \par
In addition to alterations of the equation of state in the thin, irradiated disc atmosphere, several important heating and cooling processes in this region cannot be captured by our radiative transport scheme. The optical thinness and consequential rarity of collisions result in weak coupling between the radiation field and the gas, leading to the prevalence of non-local thermodynamic equilibrium (non-LTE) effects. In those cases, line cooling by atoms, ions and molecules can dominate the cooling process \citep{Woitke2009, Sellek2024}. On the other hand, photodissociation by the stellar irradiation can destroy relevant coolants, while photoionisation, photoelectric effects and absorption of ultraviolet radiation by, for instance, polycyclic aromatic hydrocarbons (PAHs) can significantly contribute to heating of the optically thin gas \citep[e.g. ][]{Woitke2009, Wang2017}. Additionally, gas and dust temperatures tend to decouple in the upper layers of the disc, warranting a more detailed treatment of the energetic exchanges between the radiation field, the dust and the gas component \citep[e.g. ][]{Muley2023}. Since the self-consistent thermochemical modelling exceeds the intentions of this work, the temperatures achieved in the thin, irradiated disc atmosphere in our models should be regarded as approximate. The consideration of chemical processes in the thin gas and dust components, as performed by codes like \texttt{GGchem} \citep{Woitke2018} or \texttt{PRIZMO} \citep{Grassi2020} and implemented in hydrostatic disc structure models like \texttt{ProDiMo} \citep{Woitke2009}, in combination with the hydrodynamic evolution presented in our work, would result in a more complete picture of the thermodynamics of the inner disc. \par
As was the case for \citetalias{Cecil2024b}, the consideration of non-ideal magnetohydrodynamic (MHD) effects might change the evolution of the inner disc. For instance, \cite{Iwasaki2024} conclude that the region around the DZIE incorporates a transition zone, in which the radial mass transport is halted, effectively leading to an accumulation of mass, similar to the refilling of the inner disc shown in panel (b) of Fig. \ref{fig:Scurve}. Other works by \cite{Latter2012} and \cite{Faure2014} do not find a limit-cycle behaviour similar to our results in their MHD considerations of the dynamics of the DZIE. Consequently, the structure found in their models is prone to RWI \citep{Faure2015, Roberts2025}. Additionally, \cite{Roberts2025} find that the dead zone features a particular magnetic field topology as a result of magnetic flux transport near the DZIE. This field could possibly account for radial mass transport in the inner dead zone by driving a magnetic wind. Although a self-consistent treatment of non-ideal MHD in combination with the set-up of our work is computationally challenging, a reasonable compromise could consist of determining the MRI activity based on large-scale magnetic field structures, such as determined, for instance, by \cite{Steiner2025}, in combination with tabulated values of ambipolar and ohmic diffusivities, as calculated in works of \cite{Desch2015} or \cite{Williams2025}.  \par
The values of the physical and numerical parameters entering our simulations (Table \ref{tab:MREF}) have been chosen for consistency with previous work and are motivated by observational and theoretical literature. Different choices of parameter values can have an influence on the details of our simulation results, especially concerning the timescales of the quiescent and outburst phases. For instance, previous studies have indicated that the location of the inner boundary $r_\mathrm{in}$ can impact the magnitude of accretion events caused by TI to a significant extent \citep{Steiner2021, Elbakyan2025}. However, since the main focus of this work is the effect of opacity descriptions, detailed further parameter and resolution studies go beyond the intention of our investigations.\par
The calculations of the gas opacities by \cite{Malygin2014} considered a wide range of atomic and molecular lines. Recent investigations of gas mean opacities use slightly different line lists and treatments of the line pressure broadening \citep{Marigo2024}. This can lead to discrepancies between the different approaches, especially in the low-temperature regime. However, we do not expect these differences to significantly affect the qualitative results of this study. \par
The equilibrium state during quiescence, especially in the vicinity of the DZIE, is considerably delicate. Small disturbances in the density or thermal structure, such as planet-disc interactions, vortices, gravitational disturbances or changes in the effective stellar luminosity, can launch the burst cycle far sooner than our models predict. Since such disturbances are not included in our models, the state of the disc at $t=t_\mathrm{TI}$, as displayed in the first column of Fig. \ref{fig:burststructure}, has to be understood as the latest possible quiescent structure. Non-axisymmetric 2D or 3D models, allowing for the emergence and evolution of vortices or gravitationally bound clumps, as well as consideration of embedded planets or interaction with the stellar environment, could reveal different timescales of quiescence between burst cycles.

\section{Conclusion} \label{sec:conclusion}
The models presented in this work greatly expand on previously published results on the structure and radiation hydrodynamic evolution of the inner region of protoplanetary discs that include the inner, permanently MRI active region, the dust sublimation front and the transition to the dead zone. We investigated the influence of detailed frequency-dependent and mean opacity descriptions for both gas and dust on the inner disc's two-dimensional thermal structure and the emergence, evolution and consequences of periodic accretion burst cycles caused by the activation of the MRI in the inner regions of the dead zone. Additionally, we included the stellar accretion shock luminosity and analysed the stability of features produced by the burst cycles with respect to Rossby wave and Rayleigh instability. We summarise the main results as follows:
\begin{itemize}
    \item The consideration of the gas opacities as calculated by \cite{Malygin2014} has a significant influence on the structure of the inner disc. Both the dust sublimation front and the DZIE at the midplane are closer to the star than previously predicted. Consequently, the inner MRI active region during quiescent phases is restricted to smaller radii.
    \item In the FLD approximation of radiative transport, the gas opacities additionally lead to a significantly higher equilibrium temperature in the optically thin, gaseous regions.
    \item The equilibrium temperature degeneracy, inherent to the gas mean opacities, manifests itself as a sharp temperature transition, effectively separating the thermal structure of the disc into a high- and low-temperature regime and shaping the vertical structure of the MRI transition and dust sublimation front.
    \item The inclusion of the frequency-dependence of the gas opacities in the irradiation shifts the equilibrium temperature transition and adds substructure to the thermal state of the inner disc, especially in front of the inner dust rim.
    \item The dust-free regions can already become optically thick through the effect of gas opacities, especially at small wavelengths ($<0.5\,\mu\mathrm{m}$).
    \item The mechanism leading to accretion burst cycles by activation of the MRI behind the DZIE is still functional with different treatments of opacity descriptions.
    \item The influence of the gas opacities determines the initial conditions for the accretion outburst by setting the location of the DZIE. However, the thermodynamic evolution of the burst cycle itself is mainly determined by the Rosseland mean opacities of the dust. The equilibrium temperature in the viscously heated burst region on the upper branch of the S-curve is governed by a thermostat effect, where the decrease in optical thickness by sublimation of dust balances the enhanced heating by viscous energy dissipation. Only if the dust is (close to) being completely sublimated in regions with strong viscous heating do the gas opacities start to dominate the radiative cooling.
    \item Larger dust opacities lead to smaller values of both critical surface densities, $\Sigma_\mathrm{crit}^\mathrm{min}$ and $\Sigma_\mathrm{crit}^\mathrm{max}$. As a consequence, the burst cycle can be initiated in less massive inner discs. However, the maximum accretion rate and total mass accreted during the burst remain independent of the opacity description.
    \item The density spikes travelling outwards alongside the expanding MRI active region during burst cycles fulfil the conditions of both Rayleigh and Rossby wave instability. The viscously evolving density and pressure bumps placed in the inner disc by the burst can remain prone to RWI long enough to produce vortices in non-axisymmetric settings. 
    \item The burst cycles reset the density structure at the DZIE before the linear criteria for RWI are fulfilled, possibly keeping the DZIE stable during quiescent phases.
    
\end{itemize}
\noindent Our study highlights the importance of considering detailed opacity descriptions, especially for the gas, in the inner regions of protoplanetary discs where temperatures become high enough to sublimate dust. Depending on the physical processes or structures of interest, a careful treatment of gas opacities can have a crucial impact on the relevant results. This work provides an additional contribution to our understanding of the structure, evolution and planet-forming conditions of the regions of protoplanetary discs where the terrestrial planets of our Solar System and the majority of newly detected planets are located.

\begin{acknowledgements}
This research was supported by Deutsche Forschungsgemeinschaft (DFG, German Research Foundation) under grant no. 517644750. 
RK acknowledges financial support via the Heisenberg Research Grant funded by the Deutsche Forschungsgemeinschaft (DFG) under grant no.~KU 2849/9, project no.~445783058. AZ acknowledges funding from the European Union under the European Union's Horizon Europe Research and Innovation Programme 101124282 (EARLYBIRD). PS acknowledges the support of the Deutsche Forschungsgemeinschaft (DFG) through grant no. 495235860. Views and opinions expressed are those of the authors only.
\end{acknowledgements}
% ------------------------------------------------------------------
% FOOTER
% ------------------------------------------------------------------
\bibliographystyle{resources/aa}

\bibliography{literature/SETI_II}

\begin{appendix} \label{sec:appendix}

\section{Description of the dust-to-gas ratio} \label{app:d2g}

The dust sublimation temperature was parametrised analogous to \citet{Isella2005} as
\begin{equation}
    T_\mathrm{S}=2000\,\mathrm{K} \left( \frac{\rho}{1 \, \mathrm{g \, cm^{-3}}}\right)^{0.0195} \; . \label{eq:T_S}
\end{equation}
\noindent For simulations without frequency-dependent irradiation, the dust-to-gas ratio $f_\mathrm{D2G}$ was smoothed around $T_\mathrm{S}$ and centred at a radial optical depth for grey irradiation (see Sect. \ref{sec:meth_freqirr}) of $\tau_\mathrm{rad}=2/3$ according to (see \citetalias{Cecil2024b}, Sect. 4.2),
\begin{equation} \label{eq:fd2g_alt}
    \begin{split}
    f_\mathrm{D2G}= & \left \{ f_{\Delta \tau} \frac{1}{8} \left [ 1- \mathrm{tanh}\left ( \frac{T_\mathrm{g}-T_\mathrm{S}}{50 \, \mathrm{K}}  \right)\right] \left [ 1- \mathrm{tanh}(2/3-\tau_\mathrm{rad})\right ] \right \}\\  
    & \cdot \left \{ 1+ \mathrm{tanh}\left (\frac{3.0-\tau_\mathrm{rad}}{0.6} \right ) \right \} \\
    & +\left \{ f_0\frac{1}{4}\left [ 1-\mathrm{tanh} \left ( \frac{T_\mathrm{g}-T_\mathrm{S}}{50 \, \mathrm{K}}  \right) \right ] \right \} 
    \left \{ 1-\mathrm{tanh}\left (\frac{3.0-\tau_\mathrm{rad}}{0.6} \right ) \right \} \; ,
    \end{split}
\end{equation}
\noindent with $f_{\Delta \tau}=0.2/[\rho \, \kappa_\mathrm{P,d}(T_{\star,\mathrm{eff}}) \, \Delta r]-\kappa_\mathrm{P,g}(T_{\star,\mathrm{eff}},P_\mathrm{g})/\kappa_\mathrm{P,d}(T_{\star,\mathrm{eff}})$, where $T_{\star,\mathrm{eff}}$ is the effective temperature of the irradiating star (see Appendix \ref{sec:meth_acclum}), $f_0$ is the maximum dust-to-gas mass ratio and $\Delta r$ is the radial width of a computational cell. \par
When including frequency-dependent irradiation, the inner dust rim is intrinsically less sharp and better resolved. In those cases, a simple smoothing around the dust sublimation temperature suffices, 
\begin{equation}
    f_\mathrm{D2G}^{\mathrm{freqirr}}=f_0\frac{1}{2}\left [ 1-\mathrm{tanh} \left ( \frac{T_\mathrm{g}-T_\mathrm{S}}{50 \, \mathrm{K}}  \right) \right ] \; .
\end{equation}

\section{Derivation of the binned frequency-dependent irradiation flux}\label{app:F_irr}

In general, the irradiation flux can be evaluated at every distance $r$ to the host star with,
\begin{equation} \label{Eq:B1}
    F_\mathrm{irr}(r)=\int_\nu \int_{\Omega_\mathrm{s}} I (\nu,\theta', \varphi') \, e^{-\tau_\mathrm{rad}(\nu,r,\theta', \varphi')}\mathrm{cos}\theta'\, \mathrm{d\Omega_\mathrm{s}'} \mathrm{d}\nu~~~,
\end{equation}
\noindent where,
\begin{align}
    &\tau_\mathrm{rad}(\nu,r,\theta', \varphi')=\tau_0 + \int_{r_\mathrm{in}}^{r} \sigma(\nu,r,\theta', \varphi') ~\mathrm{d}r \; ,\label{eq:tau} &\\&
    \sigma(\nu,r,\theta', \varphi') =\rho_\mathrm{d}(r,\theta', \varphi')\kappa_\mathrm{d}(\nu)+\rho_\mathrm{g}(r,\theta', \varphi')\kappa_\mathrm{g}(\nu, P_\mathrm{g},T_\mathrm{g}) \; .& \label{eq:sigma}
\end{align}
\noindent Equation \ref{Eq:B1} should be understood as the flux passing through an infinitesimal surface located at a distance $r$ from the star. $I (\nu,\theta', \varphi')$ is the specific intensity, dependent on the frequency $\nu$ and the polar and azimuthal angle, $\theta'$ and $\varphi'$, of the direction of the incoming radiation. $\Omega_\mathrm{s}'$ represents the solid angle, the origin of which lies at the location of evaluation of $F_\mathrm{irr}$. $\tau_\mathrm{rad}(\nu,r,\theta', \varphi')$ is the optical depth of a specific frequency at radius $r$ in the direction given by $\theta'$ and $\varphi'$. It consists of the optical depth at the inner boundary $\tau_0$ and the optical thickness $\sigma(\nu,r,\theta', \varphi')$, composed of contributions from both gas and dust, integrated from the inner boundary at $r_\mathrm{in}$ to $r$. For the models presented in this work, we neglected the effect of the regions between the star and the inner boundary of the simulated domain on the optical depth and set $\tau_0=0$.  \par
Following the assumption that the central star radiates uniformly as a blackbody and $r\gg R_\star$, with $R_\star$ being the stellar radius, we can equate $I(\nu,\theta', \varphi')=B_\nu(\nu, T_{\star,\mathrm{eff}})$ (where $B_\nu(\nu, T_{\star,\mathrm{eff}})$ is the Planck function) and the solid angle covered by the central star is small. Hence, the optical depth can also be assumed to be uniform across the stellar disc and the irradiation flux can be expressed as,
\begin{equation}
    F_\mathrm{irr}(r)=\pi \left(\frac{R_\star}{r}\right)^2 \int_\nu  B_\nu(\nu,T_{\star,\mathrm{eff}})e^{-\tau_\mathrm{rad}(\nu, r)}\, \mathrm{d}\nu ~~~.
\end{equation}
\noindent In the numerical implementation, the spectral range of the irradiation was split into a number $n$ of discrete bins, where each bin has its respective optical depth. Therefore, we can write,
\begin{equation}
    F_\mathrm{irr}(r)=\pi \left(\frac{R_\star}{r}\right)^2 \sum_{i=1}^n e^{-\tau_\mathrm{rad}(\nu_i,r)}\int\limits_{(\nu_{i-1}+\nu_i)/2}^{(\nu_{i+1}+\nu_i)/2}B_\nu(\nu,T_{\star,\mathrm{eff}})\mathrm{d}\nu ~~~.
\end{equation}
\noindent For practical purposes, it is useful to utilise $\int_\nu B_\nu(\nu, T_{\star,\mathrm{eff}})\, \mathrm{d}\nu=\frac{\sigma_\mathrm{SB}}{\pi}T_{\star,\mathrm{eff}}^4$, with $\sigma_\mathrm{SB}$ being the Stefan-Boltzmann constant, and define the spectral weight of a frequency bin as,
\begin{equation} \label{eq:binweights}
    w_i=\frac{\int\limits_{(\nu_{i-1}+\nu_i)/2}^{(\nu_{i+1}+\nu_i)/2}B_\nu(\nu,T_{\star,\mathrm{eff}})\,\mathrm{d}\nu}{\int_\nu B_\nu(\nu,T_{\star,\mathrm{eff}})\, \mathrm{d}\nu}  ~~~~.
\end{equation}
\noindent Hence, the final expression for the irradiation flux becomes,
\begin{equation} \label{eq:app_F_irr_freqdep}
    F_\mathrm{irr}(r)= \left(\frac{R_\star}{r}\right)^2 \sigma_\mathrm{SB}T_{\star,\mathrm{eff}}^4 \sum_{i=1}^n w_ie^{-\tau_\mathrm{rad}(\nu_i,r)} ~~~.
\end{equation}
In this derivation, we neglected the dependence of the flux on the polar and azimuthal coordinates of the evaluation point for clarity. In practice, $\tau_\mathrm{rad}$ in Eq.~\ref{eq:app_F_irr_freqdep} has to be evaluated at every $\theta$ and $\varphi$ in the spherical coordinate system of the simulation, resulting in $F_\mathrm{irr}(r)=F_\mathrm{irr}(r, \theta, \varphi)$, or $F_\mathrm{irr}(r)=F_\mathrm{irr}(r,\theta)$ in axisymmetric settings. 
\section{Accretion shock luminosity feedback} \label{sec:meth_acclum}
When considering energy dissipation by the accretion process on the stellar surface, the total luminosity affecting the disc is composed of the intrinsic luminosity of the star $L_\star$ and the luminosity originating from the accretion shock $L_\mathrm{acc}$ and can be calculated with $L_\mathrm{tot}=L_\star+L_\mathrm{acc}$, where,
\begin{equation}
    L_\star=4\pi R_\star^2 \sigma_\mathrm{SB} T_\star^4 \; , ~~~L_\mathrm{acc}=\varepsilon\frac{GM_\star \dot M}{R_\star} \; .
\end{equation}
\noindent The parameter $\varepsilon$ incorporates the efficiency of the transformation of gravitational energy to luminosity and the factor resulting from the distance from which the material can be considered to be in free fall onto the star (typically the inner magnetic truncation radius of the disc) \citep{Hartmann2016}. Although $L_\mathrm{acc}$ can become larger than $L_\star$ by a factor of a few during the burst cycles, it is not expected that the accretion luminosity has a crucial influence on the dynamics of interest in our models. Hence, for simplicity, we set $\varepsilon=0.75$. \par
The total luminosity was then transformed back into an effective stellar temperature used in the evaluation of the irradiation flux in Eqs. \ref{eq:F_irr_freqdep} and \ref{eq:F_irr_nofreqirr},
\begin{equation} \label{eq:T_eff}
    T_{\star, \mathrm{eff}}=\left( \frac{L_\mathrm{tot}}{4\pi R_\star^2\sigma_\mathrm{SB}} \right )^{1/4} \; .
\end{equation}

\section{Effects of a smaller $\alpha_\mathrm{DZ}$ and different smoothing ranges around $T_\mathrm{MRI}$} \label{app:alphathings}

\begin{figure}[t]
    \centering
         \resizebox{\hsize}{!}{\includegraphics{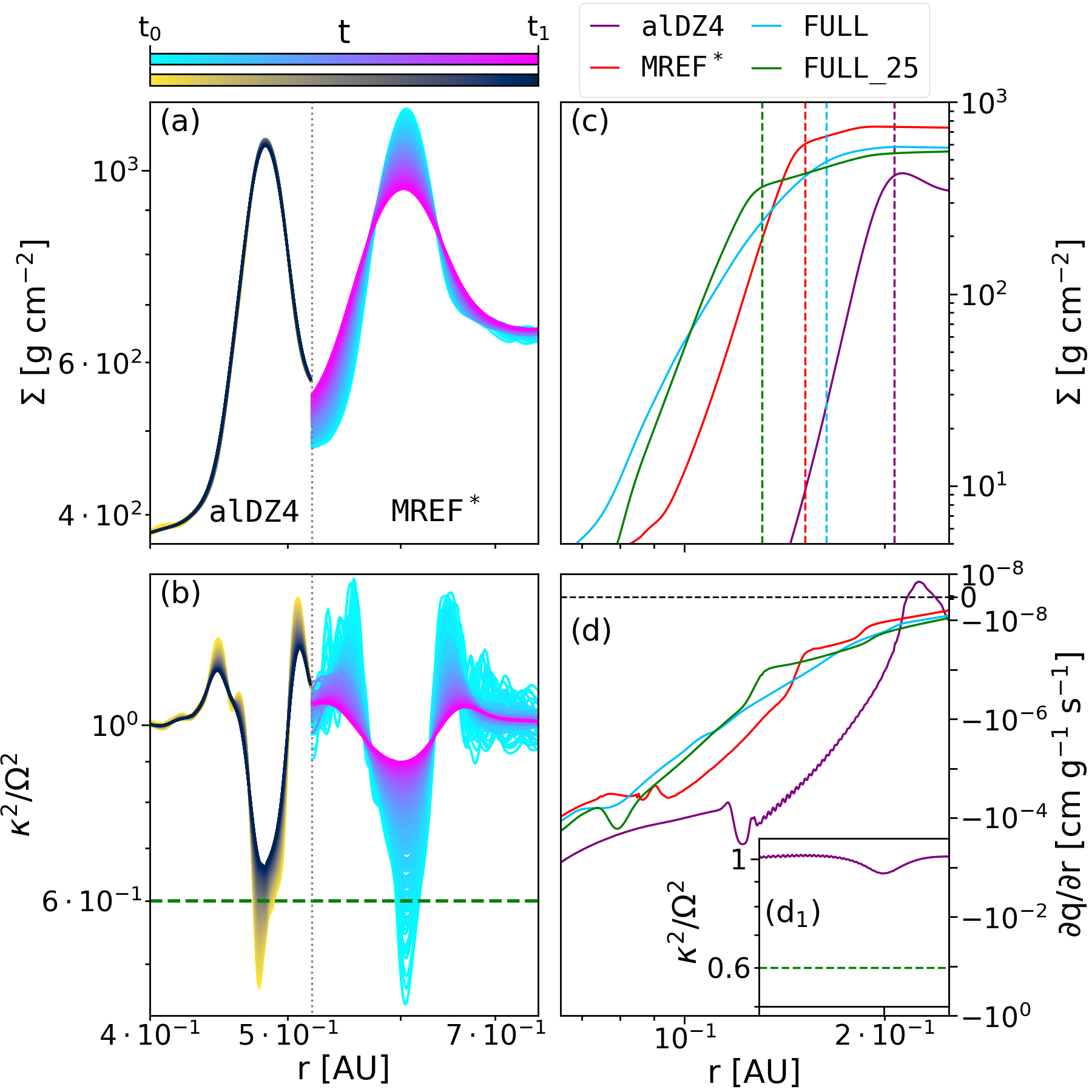}}
    \caption{Effect of different viscosity parameters on the location and evolution of pressure bumps. Panels (a) and (b) display a comparison between the models \texttt{alDZ4} and $\texttt{MREF}^*$ (separated by the grey dotted line) in terms of the evolution of the surface density and $\kappa^2/\Omega^2$ for a density bump placed by the respective outburst cycles. The depicted timeframe is the same for both models and begins at a time $t_0$ at which both models have roughly the same minimum value of $\kappa^2/\Omega^2$. The time $t_1=t_0+\Delta t$ has been chosen arbitrarily. Panels (c) and (d) show the surface density and the radial gradient of the vortensity for four different models in the radial range of the DZIE. The vertical dashed lines in panel (c) mark the positions of the inner pressure bumps in the respective models of the same colour. The inset panel ($\mathrm{d_1}$) presents $\kappa^2/\Omega^2$ for the model \texttt{alDZ4} around the position of the vortensity minimum. The green dashed lines in panels (b) and ($\mathrm{d_1}$) indicate the upper limit for the halfway-to-Rayleigh criterion.    }
    \label{fig:RWI_alphavar}
\end{figure}
For the investigation of the influence of different parameters describing the behaviour of $\alpha$, we constructed two additional models. \texttt{alDZ4} has the same configuration as $\texttt{MREF}^*$ but with a smaller value of $\alpha_\mathrm{DZ}$ of $10^{-4}$, and \texttt{FULL\_25} is equivalent to the \texttt{FULL} model, but with a sharper transition at the DZIE with $T_\Delta=25\,\mathrm{K}$ in Eq. \ref{eq_alpha}. Both models show the same phenomenology concerning the MRI-triggered outburst as the other simulations analysed in this work and \citetalias{Cecil2024b}. Panels (a) and (b) of Fig. \ref{fig:RWI_alphavar} compare the evolution of a density bump placed in the respective inner discs of the models \texttt{alDZ4} and $\texttt{MREF}^*$ by burst cycles. The timeframe shown from $t_0$ to $t_1$ corresponds to around 95 orbits at $0.5\;\mathrm{AU}$. At this location, the timeframe is equivalent to $6\cdot 10^{-4}t_\nu$ and $4.8\cdot10^{-5}t_\nu$ for $\texttt{MREF}^*$ and \texttt{alDZ4}, respectively, with $t_\nu=r^2/\nu$ being the viscous timescale. While the density bump disperses significantly during this time in $\texttt{MREF}^*$, there is no visible change in \texttt{alDZ4}. However, panel (b) shows that even for the slowly evolving \texttt{alDZ4}, the bump becomes stable to RWI after approximately 1/3 of the displayed timeframe. \par
Panel (c) compares the surface density structure and the locations of the midplane pressure maxima at the DZIE during the quiescent phases for the four models. As expected, the \texttt{FULL} model exhibits a smaller gradient and a more smoothed-out transition between the MRI active and dead zone. Consequently, the inner pressure bump is placed at almost the same location as in the $\texttt{MREF}^*$ model, although the inner dust rim is much closer to the star in \texttt{FULL}. Only if the transition range is reset to the same width as in $\texttt{MREF}^*$, the pressure bump emerges at smaller radii, as modelled in \texttt{FULL\_25}. A possible consequence of this shift in the location of the pressure bump could be that the MRI can be activated (and the burst cycle initiated) even sooner than in the \texttt{FULL} model due to the critical surface density being smaller. \par
In \texttt{alDZ4}, the DZIE and the corresponding pressure maximum move further outwards in the quiescent phase than in models with a larger $\alpha_\mathrm{DZ}$. Since $\texttt{MREF}^*$ and \texttt{alDZ4} have the same opacity description and equal values of $\alpha_\mathrm{MRI}$, $\Sigma_\mathrm{crit}^\mathrm{min}$ is also very similar since its value depends on the balance between the viscous heating of the MRI active disc and the radiative cooling. Consequently, the surface density profiles at the beginning of the quiescent phase are also approximately the same in both models. However, since the ratio $\alpha_\mathrm{MRI}/\alpha_\mathrm{DZ}$ is larger by an order of magnitude in \texttt{alDZ4}, the density in the irradiated MRI active region during quiescence is much smaller, decreasing its optical thickness, allowing the stellar radiation to penetrate deeper into the disc and to push the DZIE further outwards. \par
Another consequence of the larger $\alpha_\mathrm{MRI}/\alpha_\mathrm{DZ}$ ratio is a higher efficiency of the mass accumulation in the inner dead zone due to the exacerbated difference in angular momentum transport between the MRI active and inactive zones. This results in a stronger pressure and density bump, which could bring the DZIE closer to RWI. Panel (d) of Fig. \ref{fig:RWI_alphavar} shows the radial gradient of the vortensity of the four different models. The \texttt{alDZ4} model does indeed exhibit a vortensity minimum at the DZIE, fulfilling the Lovelace criterion. However, the inset panel ($\mathrm{d}_1$) indicates that this structure is still far away from meeting the halfway-to-Rayleigh condition and possibly remains stable. However, smaller values of $\alpha_\mathrm{DZ}$ require larger surface densities to trigger a burst cycle, which could possibly also facilitate the onset of RWI. In the snapshot of \texttt{alDZ4} shown in Fig. \ref{fig:RWI_alphavar}, the model has not reached $t=t_\mathrm{TI}$ after the quiescent phase due to computation time restrictions. Therefore, the susceptibility of the DZIE to RWI in this model is possibly underestimated. A proper assessment of these results requires simulations in non-axisymmetry.
\section{Delayed MRI saturation} \label{app:alphadelay}
To analyse the interplay between the non-instantaneous saturation of the MRI and the outburst mechanism, we adopted an approach similar to the method used by \cite{Zhu2010a}. We constructed additional simulations in which the prescription of the stress-to-pressure ratio was complemented with a delay factor $\xi$, 
\begin{equation} \label{eq:alpha_delay}
   \alpha=(\alpha_\mathrm{MRI}-\alpha_\mathrm{DZ}) \, \xi+\alpha_\mathrm{DZ} \; ,
\end{equation}
\noindent where $\xi$ follows the evolution equation,
\begin{equation} \label{eq:ksi}
    \frac{\partial \xi}{\partial t}=
    \begin{cases}
    \frac{1}{\alpha_\mathrm{MRI}-\alpha_\mathrm{DZ}} \, \left [ \mathrm{exp} \left ( 6.253 \, \frac{\Omega}{2\pi m} \alpha_\mathrm{del} \right ) -1 \right ] \vspace{0.1cm}\\
    \hspace{2.5cm} \cdot \frac{1}{2} \left [ 1+ \mathrm{tanh} \left ( \frac{0.8 \alpha_1-\alpha_\mathrm{del}}{0.25 \alpha_1} \right ) \right ] &\text{for $\xi < \xi_1$}  \\
    & \\
    -\frac{1}{\alpha_\mathrm{MRI}-\alpha_\mathrm{DZ}} \vspace{0.1cm} \\ 
    \hspace{0.3cm} \cdot \left ( \mathrm{exp} \left [ 4.57 \, \frac{\Omega}{2\pi m} (\alpha_\mathrm{MRI}+\alpha_\mathrm{DZ}  
    -\alpha_\mathrm{del}) \right ]-1 \right ) &\text{for $\xi > \xi_1$}  \\
    & \\
    0 & \text{for $\xi = \xi_1$}
     \end{cases}\; ,
\end{equation}
\noindent with,
\begin{equation} \label{eq:ksi_1}
    \xi_1 = \frac{1}{2} \left [1 - \mathrm{tanh}\left ( \frac{T_\mathrm{MRI}-T_\mathrm{g}}{T_\Delta}\right)\right] \; ,
\end{equation}
\noindent and, 
\begin{equation} \label{eq:alpha_1}
    \alpha_\mathrm{1}=(\alpha_\mathrm{MRI}-\alpha_\mathrm{DZ}) \, \xi_1+\alpha_\mathrm{DZ} \; .
\end{equation}
\noindent In this context, $\xi$ acts as a weight function that changes its value towards $\xi_1$, which is the standard smoothing factor included in Eq. \ref{eq_alpha}. $\xi$ ensures that the time it takes for $\alpha$ to transition between $\alpha_\mathrm{MRI}$ and $\alpha_\mathrm{DZ}$ after a rapid change in temperature is equal to $m$ orbits. As initial values, we set $\xi=\xi_1$. \par
\begin{figure}[t]
    \centering
         \resizebox{\hsize}{!}{\includegraphics{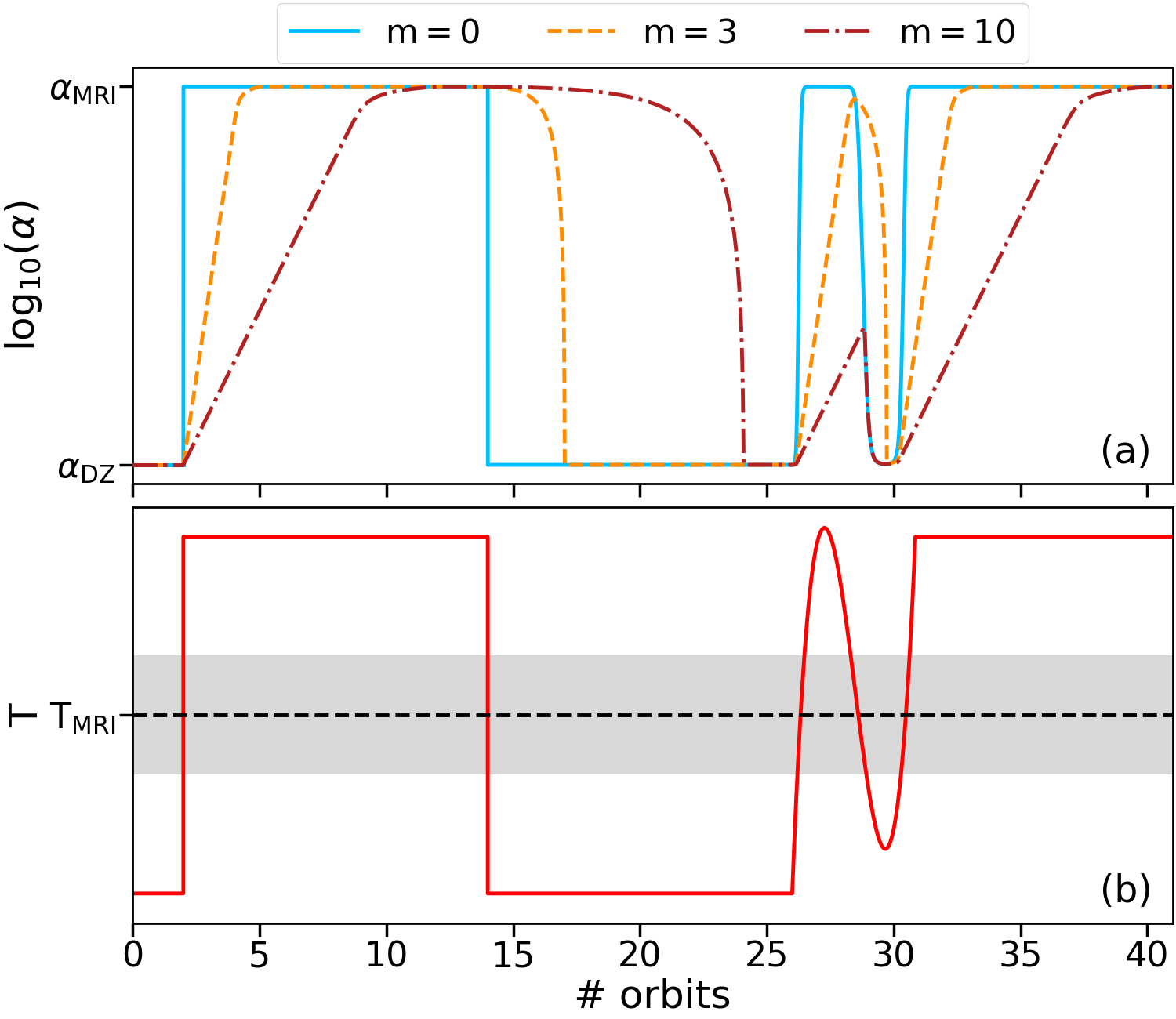}}
    \caption{Demonstration of the effect of delayed MRI-saturation. Panel (a) shows the temporal evolution of $\alpha$ between its minimum (dead zone) value $\alpha_\mathrm{DZ}$ and its maximum (fully MRI active) value $\alpha_\mathrm{MRI}$ in response to an artificial temperature evolution shown in panel (b). The dashed black line in panel (b) marks the MRI activation temperature $T_\mathrm{MRI}$, while the grey shaded area indicates the smoothing range around $T_\mathrm{MRI}$, determined by $T_\Delta$. }
    \label{fig:alphasaturation}
\end{figure}

Fig. \ref{fig:alphasaturation} visualises the behaviour of $\alpha$ in time according to Eq. \ref{eq:alpha_delay} for saturation timescales of zero, three and ten orbits. After a quick increase of the temperature above $T_\mathrm{MRI}$, $\alpha$ increases exponentially with varying slopes, dependent on the chosen timescale. The increase tapers off shortly before saturation is reached. This behaviour mimics the growth of the MRI, as investigated in e.g. \cite{Hawley1991} and \cite{Flock2010}. Rapid oscillations in temperature may remain mostly unrecognised by the viscosity in models with long MRI saturation timescales.

We investigated the consequences of the gradual saturation of the MRI by setting up two additional models, \texttt{FULL\_DEL(3)} and \texttt{FULL\_DEL(10)}, which incorporate the description of $\alpha$ via Eq. \ref{eq:alpha_delay} with saturation times of three and ten local orbits, respectively. Since the MRI transition stays approximately static during quiescence, the following analysis focuses on the burst phase. \par
\begin{figure}[t]
    \centering
         \resizebox{\hsize}{!}{\includegraphics{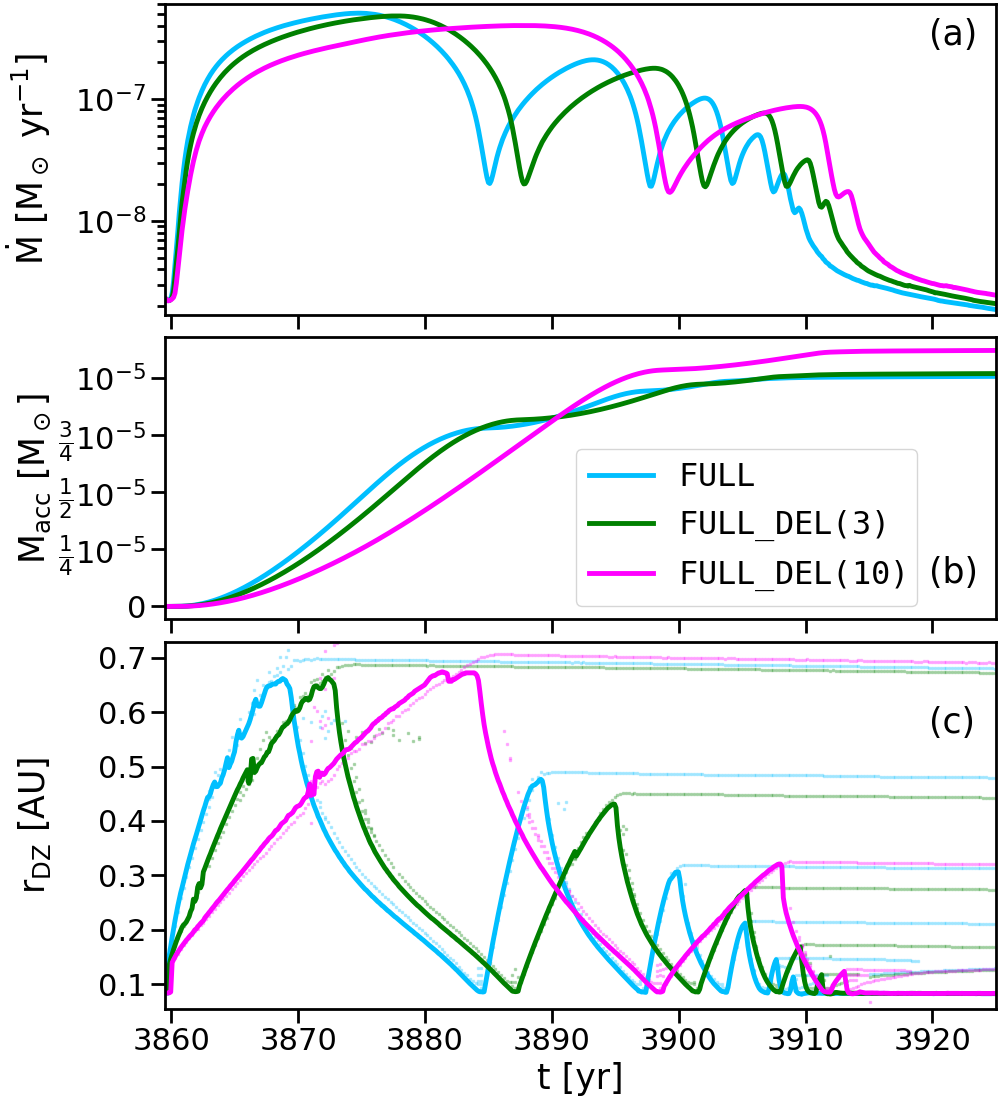}}
    \caption{Effect of different MRI saturation times during a burst cycle. Panels (a), (b) and (c) show the accretion rate, the total accreted mass and the position of the midplane DZIE, respectively, for models with instantaneous saturation (\texttt{FULL}) and saturation over the course of three (\texttt{FULL\_DEL(3)}) and ten (\texttt{FULL\_DEL(10)}) local orbits. The small dots in panel (c) mark the positions of midplane pressure maxima. }
    \label{fig:accr_rdz_alphadelay}
\end{figure}

Fig. \ref{fig:accr_rdz_alphadelay} shows the temporal evolution of the bursts occurring in models with various MRI saturation timescales. Panel (a) indicates that the \texttt{FULL\_DEL} models still involve a reflare behaviour. However, the individual flares last longer with increasing saturation time due to the advancements of both the heating and the cooling fronts being slowed down. Since the high-state of the burst phase is governed by the same $\alpha_\mathrm{MRI}$ in all cases, a larger total amount of mass is accreted onto the star per flare in the \texttt{FULL\_DEL} models. Hence, fewer reflares occur as the saturation time increases. \par
The same is also discernible in panel (b), where the cumulative total mass accreted on the star is displayed. While the differences between \texttt{FULL} and \texttt{FULL\_DEL(3)} are small, the significant delay in MRI saturation in \texttt{FULL\_DEL(10)} leads to more total mass having been accreted at the end of the burst cycle. \par
The MRI activation front at the midplane of the first flare of the accretion event reaches approximately the same distance in all three models, as illustrated in panel (c). Consequently, the outermost pressure bumps are located at the same radii as well. Due to the decreased number of reflares, the models with larger MRI saturation times produce fewer pressure maxima in the dead zone, with only two being created during the burst in the \texttt{FULL\_DEL(10)} model.

\section{Under-relaxation scheme} \label{app:underrelax}

The non-linearities introduced in the radiative transport equations by the temperature- and pressure-dependence of the opacities can lead to large oscillations in the solutions found for the total energy and, ultimately, the temperature. Since resolving these non-linear effects by significantly reducing the timestep while still maintaining the possibility to conduct long-term simulations is unfeasible, it is necessary to suppress these oscillations in time. Furthermore, our numerical scheme includes a linearization of the terms proportional to $T_\mathrm{g}^4$ appearing in Eqs. \ref{eq:rad1} and \ref{eq:rad2} \citep[adopted from][]{Commeron2011}{}{}, which is only valid if the relative change in temperature between two subsequent timesteps is small \citep[for details, we refer to][Appendix A]{Flock2013}. For these purposes, we implemented an under-relaxation scheme for the temperature and both mean opacities in the following way,
\begin{align}
    T_\mathrm{g}^{t+\delta t}=T_\mathrm{g}^t+\iota_\mathrm{T}(T_\mathrm{g, sol}^{t+\delta t}-T_\mathrm{g}^t)\; , && \iota_\mathrm{T}=\iota_0\frac{T_\mathrm{g}^t}{T_\mathrm{g,sol}^{t+\delta t}} \; , \label{eq:underrelax_T}\\
    \kappa_\mathrm{P,R}^{t+\delta t}=\kappa_\mathrm{P,R}^t+\iota_\kappa(\kappa_\mathrm{P,R,sol}^{t+\delta t}-\kappa_\mathrm{P,R}^t)\; , && \iota_\kappa=\iota_0\frac{\kappa_\mathrm{P,R}}{\kappa_\mathrm{P,R,sol}^{t+\delta t}} \; , \label{eq:underrelax_kappa}
\end{align}
\noindent where $\delta t$ is the timestep and $T_\mathrm{g,sol}^{t+\delta t}$ and $\kappa_\mathrm{P,R,sol}^{t+\delta t}$ are the solutions for the temperature and opacities for the next timestep as found by the numerical solver. $\iota_\mathrm{T, \kappa}$ can be regarded as a percentage of the offset between the current values and the newly evaluated solution. With this description, the solver algorithm is forced to adopt values that do not deviate from the current solutions to an extreme extent, while still allowing the quantities to evolve towards any value without numerical restriction. We chose the constant factor $\iota_0$ to be between 1/5 and 2/5 for all models. \par
\begin{figure}[t]
    \centering
         \resizebox{\hsize}{!}{\includegraphics{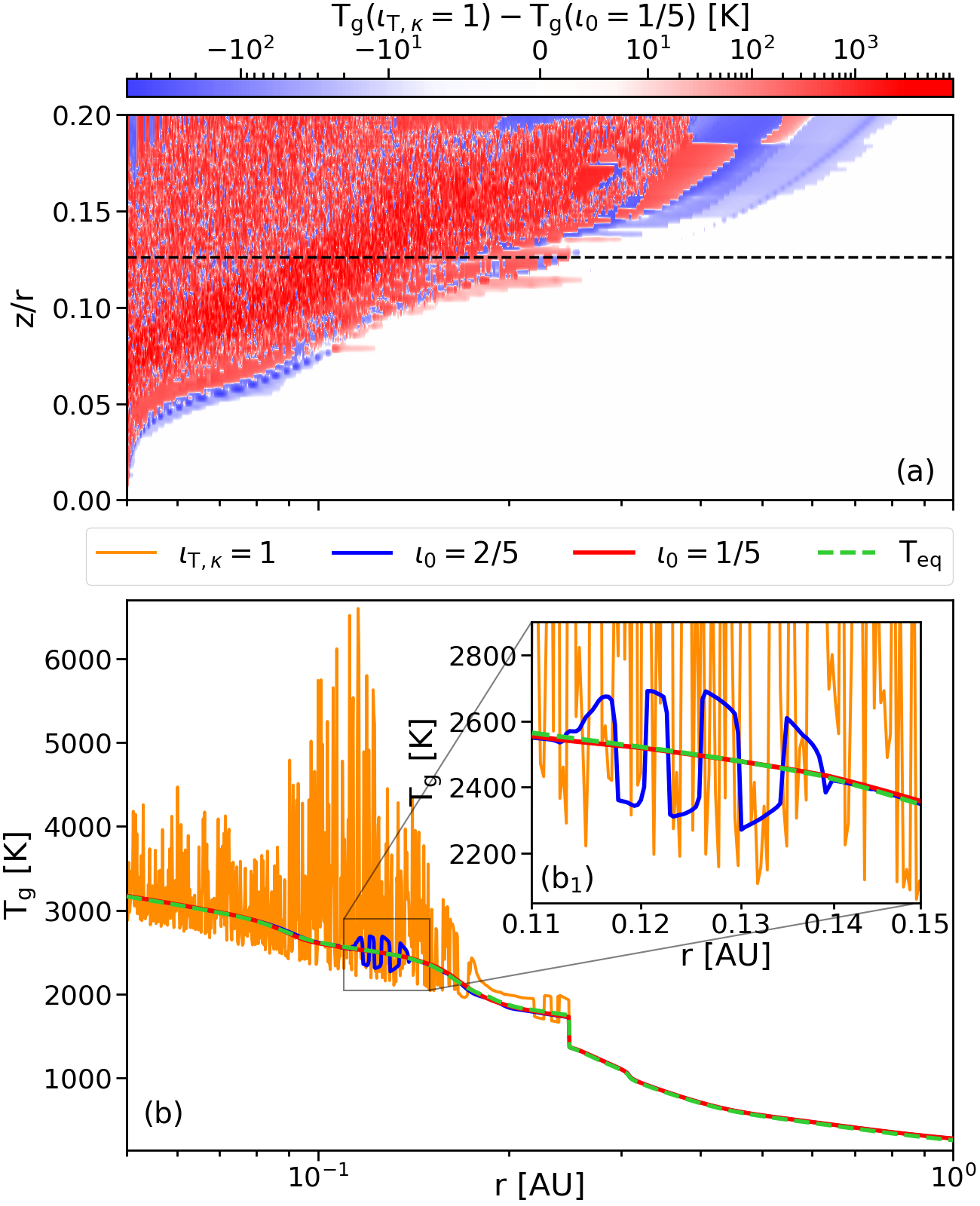}}
    \caption{Visualisation of the effect of the under-relaxation scheme. Panel (a) displays a map of the temperature residuals between states calculated with ($\iota_{0}=1/5$) and without ($\iota_{\mathrm{T},\kappa}=1$) under-relaxation for the upper hemisphere of the \texttt{FULL} model in quiescence. Radial temperature profiles for these two states have been extracted at a height of $z/r=0.125$ (black dashed line) and are shown in panel (b). The blue curve represents an additional state resulting from a larger under-relaxation parameter $\iota_0=2/5$. The profile of the equilibrium temperature evaluated at every radius with Eq. \ref{eq:T_eq} is shown as the green dashed line. Panel ($\mathrm{b_1}$) presents a magnification of the region where the more generous under-relaxation description still deviates from the equilibrium solution.  }
    \label{fig:underrelax}
\end{figure}

The oscillations only occur in the optically thin, dust-free regions of the disc. Therefore, the under-relaxation scheme is only applied in numerical cells where $f_\mathrm{D2G}\approx0$ and $\rho<7.5\cdot10^{-11}\mathrm{g\; cm^{-3}}$. The effect of this method is illustrated in Fig. \ref{fig:underrelax}. Panel (a) shows the manifestation of the strong oscillations in temperature in the optically thin regions by comparing two models with and without the effect of the under-relaxation scheme, respectively. The radial temperature profiles of these two models at a height of $z/r=0.125$ are depicted in panel (b) as the red and orange lines, respectively. The blue profile shows an intermediate state with a less restrictive under-relaxation description, where small oscillations still occur in specific regions (displayed in the inset panel $\mathrm{b_1}$). The analytical equilibrium temperature solution shown in green has been calculated with Eq. \ref{eq:T_eq}, considering all of the included terms (contribution of dust, pressure-dependence of the opacities and optical depth effects). The numerical temperature profile resulting from the under-relaxed model with $\iota_0=1/5$ is in excellent agreement with the analytical solution.

\section{Equilibrium temperature with binned frequency-dependent irradiation} \label{app:Teq}

In order to find an analytic expression for the equilibrium temperature in the optically thin, irradiated medium in the most general case, we equate the heating rate $Q^+=\rho \kappa_\mathrm{irr}F_\mathrm{irr}$ to the cooling rate $Q^-=\rho \kappa_\mathrm{P}F_\mathrm{em}$, with $\kappa_\mathrm{P}$ being the total effective Planck mean opacity (Sect. \ref{sec:opacities}), $\kappa_\mathrm{irr}$ the irradiation opacity and $F_\mathrm{em}$ the emission flux. Considering the binned frequency-dependence of the irradiation, the product $\kappa_\mathrm{irr}F_\mathrm{irr}$ has to be evaluated in every bin separately before summing over all bins. Using the spectral weights of the frequency bins introduced in Eq. \ref{eq:binweights} and considering the attenuation of the irradiation up to the evaluation radius $r$, the heating rate is expressed as,
\begin{equation}
    Q^+(r)=\rho \sum_{i=1}^n\kappa(\nu_i, T_\mathrm{g}, P_\mathrm{g})w(\nu_i)F_0e^{-\tau_\mathrm{rad}(\nu_i,r)}  \;,
\end{equation}
\noindent where $\kappa(\nu_i,T_\mathrm{g},P_\mathrm{g})=\kappa_\mathrm{g}(\nu_i,T_\mathrm{g},P_\mathrm{g})+f_\mathrm{D2G}\kappa_\mathrm{d}(\nu_i)$ takes the contribution of both gas and dust into account. With $F_0=\left(R_\star/r\right)^2\sigma_\mathrm{SB}T_\star^4$, we define the irradiation opacity $\kappa_\mathrm{irr}(T_\mathrm{g},P_\mathrm{g})$ as,
\begin{equation}
    \kappa_\mathrm{irr}(T_\mathrm{g},P_\mathrm{g})=\sum_{i=1}^n\kappa(\nu_i, T_\mathrm{eq}, P_\mathrm{g})w(\nu_i)e^{-\tau_\mathrm{rad}(\nu_i,r)} \;.
\end{equation}
In the case of grey irradiation, the irradiation opacity can be simplified to $\kappa_\mathrm{irr}(T_\mathrm{g}, P_\mathrm{g})=\kappa_\mathrm{P}(T_\star, P_\mathrm{g})$.
Identifying the volumetric emission rate of the medium, assuming black-body behaviour and no re-absorption of emitted radiation, with $F_\mathrm{em}=a_\mathrm{R}cT_\mathrm{g}^4=4\sigma_\mathrm{SB}T_\mathrm{g}^4$, setting $Q^+=Q^-$ and solving for $T_\mathrm{g}=T_\mathrm{eq}$, we arrive at the final expression for the equilibrium temperature \citep[analogous to][]{Chiang1997},
\begin{equation} \label{eq:T_eq}
    T_\mathrm{eq}=\left( \frac{\kappa_\mathrm{irr}(T_\mathrm{eq},P_\mathrm{g})}{\kappa_\mathrm{P}(T_\mathrm{eq}, P_\mathrm{g})}\right ) ^{1/4}\left(\frac{R_\star}{2r} \right )^{1/2}T_\star ~~~~.
\end{equation}

\section{Comments on timescales} \label{app:timescales}
As described in Sect. \ref{sec:dis_opacnecessity}, the inclusion of mean and frequency-dependent dust and gas opacities significantly increases the computational cost of the models. While it was possible to include the entire quiescent phase in addition to the two burst phases in the $\texttt{MREF}^*$ model, the simulation of the long quiescent phase had to be artificially accelerated in \texttt{FULL}. For that purpose, we let the model relax into the quiescent state after the initial burst cycle before we accelerated the computation by switching to a simple Euler solver (in contrast to the second-order Runge-Kutta method normally used for all models) and shortening the viscous evolution timescale by increasing both\footnote{The position of the DZIE and the efficiency of mass accumulation in the inner dead zone is governed by the ratio $\alpha_\mathrm{MRI}/\alpha_\mathrm{DZ}$, which remained unchanged.} $\alpha_\mathrm{MRI}$ and $\alpha_\mathrm{DZ}$ by a factor of 1.3. Since the disc structure is not subject to any rapid changes during quiescence, the simpler solver algorithm still captures the relevant processes to sufficient accuracy.\par
We used the slope of the accretion rate evolution depicted in panel (b) of Fig. \ref{fig:radius_time} as a diagnostic for when the relaxation after the initial burst cycle has been completed. Taking the $\texttt{MREF}^*$ model as a template, we switched on the acceleration as soon as the slope of the accretion rate evolution became constant in time. The acceleration was stopped when the model entered the next burst phase. At this point, we reverted to our standard computational method, decreased the $\alpha$ parameters back to their normal values and let the model relax again until the accretion rate evolution adopted the same slope as had been established before the acceleration. The next burst cycle then emerged soon after this relaxation phase was completed at a critical accretion rate. The dashed line in panel (b) of Fig. \ref{fig:radius_time} shows the interpolated accretion rate, having the slope of the relaxed quiescent model, during the timeframe of the acceleration. \par
This method of shortening the computational time might alter the actual timescale of the quiescent phase between burst cycles. However, as elaborated on in Sect.~\ref{sec:dis_limits}, the duration of the quiescent phase may also be subject to other processes or interactions that are not considered in our models but could disturb the equilibrium state in the inner dead zone and ignite bursts at different stages. Therefore, predicting the quiescent timescale is a complex endeavour, especially in non-isolated systems.

\section{Relevant expressions for RWI conditions} \label{app:RWI_expressions}
The vortensity $q$ is given as the fraction between the vorticity and the surface density,
\begin{equation}
    q=\frac{(\mathrm{rot\;\vec{v})_\mathrm{z}}}{\Sigma}=\frac{\kappa_\mathrm{ef}^2}{2\Omega\Sigma} \; ,
\end{equation}
with the squared epicyclic frequency being defined as,
\begin{equation}
    \kappa_\mathrm{ef}^2=2\Omega (\mathrm{rot\;\vec{v})_\mathrm{z}}=\frac{2\Omega}{r}\left ( v_\varphi +r \frac{\mathrm{d}v_\varphi}{\mathrm{d}r}\right) \;.
\end{equation}
The Gaussian profile fitted to the density bumps in our models is given as (analogous to \cite{Ono2016}),
\begin{equation}
    \Sigma=\Sigma_\mathrm{bg}\left [1+A\;\mathrm{exp}\left(-\frac{(r-r_\mathrm{bump})^2}{2W^2} \right) \right] \; ,
\end{equation}
with $\Sigma_\mathrm{bg}=\Sigma_0r^b$ being the background surface density profile (without bumps), where $\Sigma_0$ and $b$ are the background surface density at the bump location $r_\mathrm{bump}$ and the slope of the background profile, respectively.

\end{appendix}

\end{document}